\documentclass[twocolumn]{aastex631}

\graphicspath{{./}{figures/}}

\begin{document}

\title{Molecular outflow in the reionization-epoch quasar J2054-0005 revealed by OH 119 \(\micron\) observations}

\correspondingauthor{Dragan Salak}
\email{dragan@oia.hokudai.ac.jp}

\author[0000-0002-3848-1757]{Dragan Salak}
\affiliation{Institute for the Advancement of Higher Education, Hokkaido University, Kita 17 Nishi 8, Kita-ku, Sapporo, Hokkaido 060-0817, Japan}
\affiliation{Department of Cosmosciences, Graduate School of Science, Hokkaido University, Kita 10 Nishi 8, Kita-ku, Sapporo, Hokkaido 060-0810, Japan}

\author{Takuya Hashimoto}
\affiliation{Tomonaga Center for the History of the Universe (TCHoU), Faculty of Pure and Applied Science, University of Tsukuba, Ibaraki, 305-8571, Japan}

\author{Akio K. Inoue}
\affiliation{Department of Physics, School of Advanced Science and Engineering, Faculty of Science and Engineering, Waseda University, 3-4-1, Okubo, Shinjuku, Tokyo 169-8555, Japan}
\affiliation{Waseda Research Institute for Science and Engineering, Faculty of Science and Engineering, Waseda University, 3-4-1 Okubo, Shinjuku, Tokyo 169-8555, Japan}

\author{Tom J. L. C. Bakx}
\affiliation{Department of Space, Earth, \& Environment, Chalmers University of Technology, Chalmersplatsen 4 412 96 Gothenburg, Sweden}

\author[0000-0001-5341-2162]{Darko Donevski}
\affiliation{National Centre for Nuclear Research (NCBJ), Pasteura 7, 02-093 Warsaw, Poland}
\affiliation{SISSA, ISAS, Via Bonomea 265, Trieste I-34136, Italy}
\affiliation{IFPU, Institute for fundamental physics of the Universe, Via Beirut 2, I-34014 Trieste, Italy}

\author{Yoichi Tamura}
\affiliation{Division of Particle and Astrophysical Science, Graduate School of Science, Nagoya University, Aichi 464-8602, Japan}

\author{Yuma Sugahara}
\affiliation{Waseda Research Institute for Science and Engineering, Faculty of Science and Engineering, Waseda University, 3-4-1 Okubo, Shinjuku, Tokyo 169-8555, Japan}
\affiliation{National Astronomical Observatory of Japan, 2-21-1 Osawa, Mitaka, Tokyo 181-8588, Japan}

\author{Nario Kuno}
\affiliation{Tomonaga Center for the History of the Universe (TCHoU), Faculty of Pure and Applied Science, University of Tsukuba, Ibaraki, 305-8571, Japan}

\author{Yusuke Miyamoto}
\affiliation{Department of Electrical, Electronic and Computer Engineering, Fukui University of Technology, 3-6-1 Gakuen, Fukui, Fukui 910-8505, Japan}

\author{Seiji Fujimoto}
\affiliation{Department of Astronomy, The University of Texas at Austin, Austin, TX 78712, USA}

\author{Suphakorn Suphapolthaworn}
\affiliation{Department of Cosmosciences, Graduate School of Science, Hokkaido University, Kita 10 Nishi 8, Kita-ku, Sapporo, Hokkaido 060-0810, Japan}

\begin{abstract}

Molecular outflows are expected to play a key role in galaxy evolution at high redshift. To study the impact of outflows on star formation at the epoch of reionization, we performed sensitive ALMA observations of OH 119 \(\micron\) toward J2054-0005, a luminous quasar at \(z=6.04\). The OH line is detected and exhibits a P-Cygni profile that can be fitted with a broad blue-shifted absorption component, providing unambiguous evidence of an outflow, and an emission component at near-systemic velocity. The mean and terminal outflow velocities are estimated to be \(v_\mathrm{out}\approx670~\mathrm{km~s}^{-1}\) and \(1500~\mathrm{km~s}^{-1}\), respectively, making the molecular outflow in this quasar one of the fastest at the epoch of reionization. The OH line is marginally spatially resolved for the first time in a quasar at \(z>6\), revealing that the outflow extends over the central 2 kpc region. The mass outflow rate is comparable to the star formation rate (\(\dot{M}_\mathrm{out}/\mathrm{SFR}\sim2\)), indicating rapid (\(\sim10^7~\mathrm{yr}\)) quenching of star formation. The mass outflow rate in a sample of star-forming galaxies and quasars at \(4<z<6.4\) exhibits a positive correlation with the total infrared luminosity, although the scatter is large. Owing to the high outflow velocity, a large fraction (up to \(\sim50\%\)) of the outflowing molecular gas may be able to escape from the host galaxy into the intergalactic medium.

\end{abstract}


%%%%%%%%%%%%%%%%%%
\section{Introduction}\label{sec:intro}
%%%%%%%%%%%%%%%%%%

Quasar feedback is one of the fundamental processes that regulate galaxy evolution. Galaxies acquire gas via accretion from the intergalactic medium (IGM) and through merging, and lose a fraction of their gas via galactic outflows powered by feedback from starbursts and/or active galactic nuclei (AGNs). The baryon cycle is believed to regulate how much molecular gas is available for star formation and the growth of the central supermassive black holes (SMBHs) (e.g., \citealt{MQT05,Vei05,Vei20,HQM12,ZK14,Tum17}).

Recent observations have revealed the existence of massive (stellar mass \(M_\star\sim10^{11}M_\sun\)) galaxies with diminished star formation activity already at \(z\gtrsim6\), indicating that these objects experienced a vigorous starburst episode at an earlier epoch followed by quenching (e.g., \citealt{Str14,Gla17,GBC19,Mer19,Car20,Car23,For20,Val20,San21,Lab23,Los23,Nan23}). When and how this quenching occurred is still debated, but quasar feedback is one of the possible mechanisms considered as a leading internal process to explain the rapid (\(<1~\mathrm{Gyr}\)) inside-out quenching of star formation in massive galaxies \citep{Tac15,Bar18,Cos18,Spi19,Bis21,MF23}. Quasars at \(z\sim6\), powered by AGN and star formation, are thus believed to be an important evolutionary phase in massive galaxy evolution (e.g., \citealt{CW13,Cas14,Lap18}). To understand how massive galaxies evolved, it is important to reveal the physical conditions of the interstellar medium (ISM) in quasars at the epoch of reionization (EoR; \(6\lesssim z\lesssim20\)), and this has been at the focus of recent research (e.g., \citealt{Ven17,Ven18,Wal18,Has19,Nov19,Li20a,Li20b,Nee21,Mey22,Pen22,Dec23}). Many of these quasars exhibit high far-infrared luminosities (\(L_\mathrm{FIR}\gtrsim10^{13}~L_\sun\)) that indicate dust heating by extreme star formation and AGN radiation (e.g., \citealt{Wal09,Wan13,Lei14,Ven18,Ven20}). Extrapolating from the known properties of local galaxies \citep{Fle19,Lut20,RB20}, it is expected that energy released in such nuclear activity is sufficient to generate galactic outflows. Since molecular gas is the primary fuel for star formation and SMBH growth, it is important to investigate the molecular phase, which has been largely untraced, at the EoR.

While AGN-driven outflows have been observed extensively at low/moderate redshifts, and the relations between the outflows and the physical properties of host galaxies investigated (e.g., \citealt{Rup05,RTD21,Fer10,Vei13,Cic14,Fio17,Gow18,FS19,Lut20}), our understanding of quasar feedback in the early Universe has been limited. This is because detecting outflows from the emission lines of standard tracers such as CO relies on identification of broad wings in the spectra that are generally much weaker than the main component of the line profile and requires a high signal-to-noise ratio (e.g., \citealt{Cic14}). Most high-\(z\) outflow studies are based on searches of broad wings in the emission spectra of [\ion{C}{2}] 158 \(\micron\) and CO lines (e.g., \citealt{Dec18,Nov20}) and extended halos of cold gas \citep{Fuj19,Fuj20}, but unambiguous detections of outflows (distinguished from inflows) using these lines are still rare at the EoR and beyond \citep{Izu21}, making it difficult to evaluate the quasar-driven feedback.

With a relative abundance of \([\mathrm{OH}]/[\mathrm{H_2}]\sim1\times10^{-7}\) to \(\sim5\times10^{-6}\) in nearby galaxies, hydroxyl (OH) is one of the important molecular species in the ISM \citep{Wei63,Sto81,Goi02,Goi06,Ngu18}. Recent observations have shown that the OH \(^{2}\Pi_{3/2}\) \(J=5/2\leftarrow3/2\) absorption line at \(\lambda_\mathrm{rest}=119~\micron\) has proved to be a robust tracer of outflows in nearby ultraluminous infrared galaxies (ULIRGs) including AGNs \citep{Fis10,Vei13,GA14,GA17,Spo13,Cal16,Sto16,Run20}. The line is a doublet (rest wavelengths \(119.23~\micron\) and \(119.44~\micron\)) with near-equal intensities due to the \(\Lambda\)-doubling of rotational energy levels. Each of these is further split due to hyperfine structure, although these usually remain unresolved in extragalactic observations.

The 119 \(\micron\) doublet can unambiguously reveal the presence of cold molecular outflows and/or inflows through its P-Cygni profile (e.g., \citealt{Vei13,HC20}). Since the energy required for the excitation of OH \(^{2}\Pi_{3/2}\) from the rotational ground state \(J=3/2\) to the state \(J=5/2\) is \(E/k\approx120~\mathrm{K}\), where \(k\) is the Boltzmann constant, cold gas (\(\lesssim100~\mathrm{K}\)) is observed in absorption against a bright continuum source. On the other hand, the gas density required to thermalize the rotational transitions of OH is very high (\(n_\mathrm{H_2}\gtrsim10^9~\mathrm{cm^{-3}}\)), so the transition can be observed in  \(J=5/2\rightarrow3/2\) emission in environments where molecular gas is highly excited (dense warm gas, either through shocks, or because it is exposed to strong far-infrared continuum radiation), such as those in AGNs \citep{Vei13}.

At high redshift, previous works have showed that OH outflows can readily be detected in strongly lensed, dusty star-forming galaxies up to \(z\approx5\) \citep{Geo14,Spi18,Spi20a,Spi20b}; there are also reports of two OH detections in quasars at \(z\approx6\) \citep{But23}, and one tentative \citep{HC20}. Interestingly, the results in \citet{Spi20a} suggest that OH 119 \(\micron\) may be a more reliable tracer of line-of-sight outflows at high \(z\) than [\ion{C}{2}] 158 \(\micron\) and CO lines. It is therefore of great interest, and motivation of this work, to investigate whether the 119 \(\micron\) line can provide a good probe of outflows at the EoR.

To search for molecular outflows in EoR quasars, we observed OH 119 \(\micron\) toward J2054-0005 using the Atacama Large Millimeter/submillimeter Array (ALMA). The quasar was discovered from the Sloan Digital Sky Survey (SDSS) data \citep{Jia08}, and later detected by ALMA in continuum as well as [\ion{C}{2}] 158 \(\micron\), [\ion{O}{3}] 88 \(\micron\), and CO lines \citep{Wan10,Wan13,Has19,Ven20}. The measurements of these lines have determined its redshift to be \(z=6.0391\pm0.0002\). The bolometric luminosity of the source is \(L_\mathrm{bol}\approx1.2\times10^{47}~\mathrm{erg~s^{-1}}\), corresponding to \(3.2\times10^{13}~L_\sun\) \citep{Far22}. The total IR luminosity of \(L_\mathrm{IR}\approx1.3\times10^{13}~L_\sun\) suggests a star formation rate (SFR) of \(\approx1900~M_\sun~\mathrm{yr}^{-1}\) \citep{Has19}, where a \citet{Kro01} initial mass function (IMF) is assumed, although this is an upper limit because of possible AGN contribution to dust heating \citep{Sch15,DiM23}. As discussed in Section \ref{subsec:agn} below, the contribution of AGN to \(L_\mathrm{IR}\) is estimated to be \(\approx59\%\) in this source, yielding a lower star formation rate of \(\mathrm{SFR}\approx800~M_\sun~\mathrm{yr}^{-1}\). However, despite the intense star formation and the presence of an AGN, neither [\ion{C}{2}] 158 \(\micron\), [\ion{O}{3}] 88 \(\micron\), nor CO lines have revealed outflows in previous observations. Is there no outflow, or is it difficult to detect it with these tracers? Establishing a reliable tracer of molecular outflows is essential for future studies of galaxies at highest redshifts.

The paper is organized as follows. In Section \ref{sec:obs}, we describe the ALMA observations and data reduction. The resulting continuum image and OH 119 \(\micron\) spectrum is presented in Section \ref{sec:res}. This is followed by an analysis of the OH gas outflow in Section \ref{sec:out}, discussion on the outflow's driving mechanism and imprint on galaxy evolution in Section \ref{sec:dis}, and a summary in Section \ref{sec:sum}.

We adopt an \(\Lambda\)CDM cosmology with parameters \(H_0=70~\mathrm{km~s^{-1}~Mpc^{-1}}\), \(\Omega_\mathrm{m}=0.3\), \(\Omega_\mathrm{\Lambda}=0.7\), and flat geometry, consistent with the measurements reported in \citet{PC20}.

%%%%%%%%%%%%%%%%%%%%%%%%%%%%
\section{Observations and data reduction}\label{sec:obs}
%%%%%%%%%%%%%%%%%%%%%%%%%%%%

The observations were conducted between August 4 and 12 in 2022 during ALMA cycle 8. The antennas of the 12 m array observed toward a single field centered at \((\alpha,\delta)_\mathrm{ICRS}=(\mathrm{20^h54^m06\fs503,-00\arcdeg05\arcmin14\farcs43})\), which corresponds to the central position of the ALMA \(87~\micron\) continuum \citep{Has19}. In most observing runs, 44 antennas were used, but the number ranged from 41 to 46. The array was in configuration C-5 with baselines from 15 m to 1301 m.

The Band 7 receivers were tuned to cover the OH doublet at an observing frequency for the adopted redshift \(z=6.0391\). Two spectral windows (upper sideband; USB) were centered at the observing frequencies 356.315 GHz and 358.090 GHz for the line observations. Since the bandwidth of each of them is 1.875 GHz, this setup makes the two spectral windows next to each other with an overlap of \(100~\mathrm{MHz}\). The 119.23 \(\micron\) line of the doublet was set to lie in this overlap region, whereas the 119.44 \(\micron\) line is separated in velocity by \(\approx521~\mathrm{km~s^{-1}}\). With a frequency resolution of 7.813 MHz, the achieved velocity resolution (average over the bandwidth) is \(6.56~\mathrm{km~s^{-1}}\). To improve the signal-to-noise ratio for the analysis, we smoothed the data cubes to a resolution of \(35~\mathrm{km~s^{-1}}\). There are 44 velocity channels in each window, with one perfectly overlapping channel, and the total effective velocity coverage of the two adjacent windows is \(3045~\mathrm{km~s^{-1}}\) with the 119.23 \(\micron\) line at the center.

The other two spectral windows (lower sideband; LSB) were dedicated to continuum observations. The central observing frequencies were 344.2 GHz and 346.1 GHz, and the bandwidth of each of them was 2 GHz.

The scheduling block was executed 9 times. J2253+1608 (8 data sets) and J1924-2914 (1 data set) were observed for the flux and bandpass calibration, whereas J2101+0341 (all data) was observed for the phase calibration. The total on-source time was 7.3 hours, whereas the total time including calibrator observations and other overheads was 12.7 hours. The uncertainties in the paper are only statistical errors; the absolute flux accuracy in Band 7 is reported to be \(10\%\) \citep{Bra21}. 

The data were reduced using the Common Astronomy Software Applications (CASA) package \citep{TCT22}. Basic calibration was performed with a CASA pipeline resulting in 9 calibrated measurement sets. The calibrated data were then combined and imaged using the CASA task \texttt{tclean}.

The continuum image was reconstructed using the line-free LSB spectral windows in the multi-frequency synthesis mode with the \texttt{hogbom} deconvolver and standard gridding. The weighting was set to \texttt{briggs} with the robust parameter equal to 0.5 (compromise between resolution and sensitivity). To conduct unbiased mask-based image reconstruction, we employed the \texttt{auto-multitresh} tool \citep{Kep20}. We also tried \texttt{tclean} in interactive mode, but there was no obvious difference in the result, so we adopted the automatically-created image. The threshold for iterations was set to \(2~\sigma\), where \(\sigma\) was calculated using the CASA task \texttt{imstat} on a first-generation clean image masked for the central region where the source is located. The rms sensitivity in the final continuum image is \(\sigma=13~\mathrm{\mu Jy~beam^{-1}}\). The synthesized beam size (full-width half maximum; FWHM) is \((b_\mathrm{maj},b_\mathrm{min})=(0\farcs205,0\farcs176)\), corresponding to \(\approx1~\mathrm{kpc}\). For the adopted cosmological parameters, \(1\arcsec\) is equivalent to \(5.689~\mathrm{kpc}\) and the luminosity distance to the source is \(D_L=58.1465~\mathrm{Gpc}\). These are the highest spatial resolution observations of OH toward a quasar at \(z>6\).

The OH image was reconstructed from USB spectral windows. The visibilities of two spectral windows were processed by \texttt{tclean} separately and continuum was not subtracted. The deconvolver and weighting setups were the same as for the continuum, and the \texttt{auto-multitresh} tool was used. The mean sensitivity over two spectral windows is \(\sigma=0.13~\mathrm{mJy~beam^{-1}}\) in a channel of \(35~\mathrm{km~s}^{-1}\), and the synthesized beam size is \((b_\mathrm{maj},b_\mathrm{min})=(0\farcs204,0\farcs174)\). The two spectral windows were merged using the task \texttt{imageconcat}.

The velocity is expressed in radio definition with respect to the rest frame of the source (\(z=6.0391\)). The frequency that corresponds to zero velocity is \(\nu=\nu_\mathrm{rest}(1+z)^{-1}=357.193\pm0.010~\mathrm{GHz}\), where \(\nu_\mathrm{rest}=2514.31640360~\mathrm{GHz}\) is the rest frequency of the OH \(^{2}\Pi_{3/2}\) \(J=3/2\mathrm{-}5/2\), \(F=3^{-}\mathrm{-}2^{+}\) transition,\footnote{All spectral line frequencies are taken from the database Splatalogue (https://splatalogue.online/).} where \(F\) is the quantum number for the total angular momentum of the molecule (including nuclear spin), and \(``+"\) and \(``-"\) denote the \(\Lambda\)-doubling of energy levels.

The final images were corrected for the primary beam attenuation. The basic parameters of the resulting images are summarized in Table \ref{tab:imp}.

\begin{table}[ht!]
\begin{center}
\caption{Image Parameters\label{tab:imp}}
\begin{tabular}{lcc}
\tableline\tableline
Parameter & Continuum & OH Cube \\
\tableline
FWHM \(b_\mathrm{maj}\) (arcsec) & 0.205 & 0.204  \\
FWHM \(b_\mathrm{min}\) (arcsec) & 0.176 & 0.174  \\
Beam position angle (degree) & 81.7 & 75.0  \\
Total bandwidth (GHz) & 4.0 & 3.65 \\
Velocity resolution (km s\(^{-1}\)) & ... & 35 \\
Sensitivity \(\sigma\) (\(\mathrm{mJy~beam^{-1}}\)) & 0.013 & 0.13 \\
\tableline
\end{tabular}
\end{center}
\tablecomments{The beam size and sensitivity of OH are mean values from two spectral windows. The sensitivity of the OH cube is in a channel of \(35~\mathrm{km~s}^{-1}\).}
\end{table}

%%%%%%%%%%%%%%%%
\section{Results}\label{sec:res}
%%%%%%%%%%%%%%%%

We begin this section by a presentation of the continuum image. It is followed by a description of the OH spectrum and derivation of its basic properties.

%%%%%
\subsection{123-\micron~continuum emission}\label{subsec:con}

\begin{figure*}[ht!]
\epsscale{1.15}
\plotone{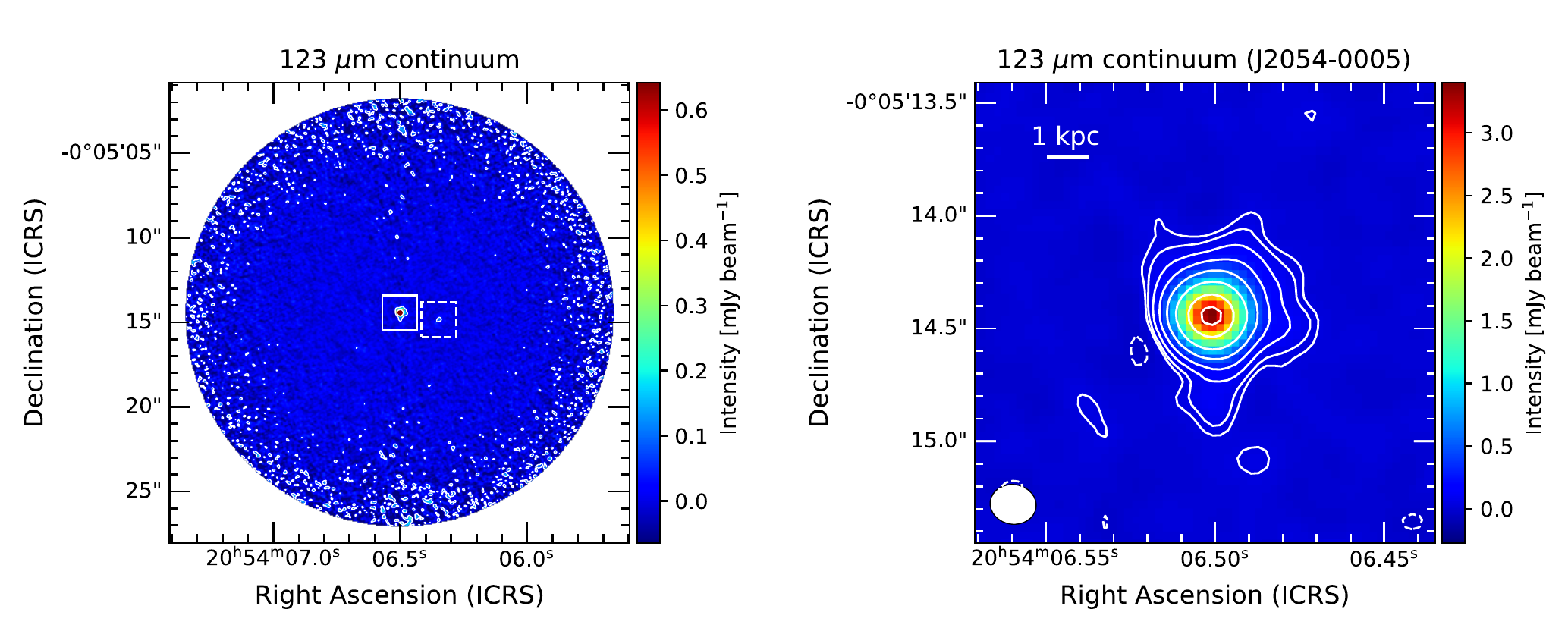}
\caption{\emph{Left.} The \(123~\micron\) continuum image. The solid rectangle indicates the position of J2054-0005; a close-up view is shown in the right panel. The dashed rectangle indicates the position of a projected companion (see Figure \ref{fig:comp}). To guide the eye, the contours are plotted at \((5,50)\times\sigma\), where \(\sigma=1.28382\times10^{-5}~\mathrm{Jy~beam^{-1}}\). The image is corrected for the primary beam attenuation, so the noise is increased at the edges. \emph{Right.} Close-up view of the continuum in J2054-0005. The contours are plotted at \((-3,3,5,10,20,40,80,160,240)\times\sigma\). The beam size is shown at the bottom left as a filled ellipse.\label{fig:cont}}
\end{figure*}

The continuum emission (\(\lambda_\mathrm{rest}=123~\micron\)), extracted from the LSB, is detected toward the quasar with a high signal-to-noise ratio of \(\mathrm{S/N}=260\) (Figure \ref{fig:cont}). The emission is spatially resolved, although strongly concentrated in the center. The flux density in the region within a radius of \(0\farcs5\) of the brightest pixel is \(S_\nu(r<0\farcs5)=5.723\pm0.009~\mathrm{mJy}\).
We estimated the peak coordinates by two-dimensional gaussian fitting of a region of radius \(0\farcs5\) centered at the brightest pixel. Using CASA's \texttt{imfit}, we obtained \((\alpha,\delta)_\mathrm{ICRS}=(20^\mathrm{h}54^\mathrm{m}06\fs501,-0\arcdeg05\arcmin14\farcs44)\). Since \(\mathrm{S/N}\) is very high, the positional uncertainty is determined by the absolute astrometric accuracy of ALMA observations in Band 7, which is \(5\%\) of the synthesized beam size (\(\approx10~\mathrm{mas}\)). By comparison, the peak of the SDSS optical \(z\)-band image is at \((\alpha,\delta)_\mathrm{ICRS}=(20^\mathrm{h}54^\mathrm{m}06\fs486,-0\arcdeg05\arcmin14\farcs50)\) with an uncertainty of \(\approx470~\mathrm{mas}\), and the peak position of the [\ion{O}{3}] 88 {\micron} is at \((\alpha,\delta)_\mathrm{ICRS}=(20^\mathrm{h}54^\mathrm{m}06\fs503,-0\arcdeg05\arcmin14\farcs48)\) with an uncertainty of \(\approx48~\mathrm{mas}\) \citep{Has19}. The \(123-\micron\) dust continuum, ionized gas traced by [\ion{O}{3}] 88 {\micron}, and \(z\)-band optical peak positions are thus in agreement within their respective uncertainties. The peak intensity obtained from the gaussian fitting is \(3.308\pm0.014~\mathrm{mJy~beam^{-1}}\), and the size (FWHM) of the central region where the emission is concentrated, deconvolved from the beam, is estimated to be \((d_\mathrm{maj},d_\mathrm{min})=(0\farcs1567\pm0\farcs0017,0\farcs1321\pm0\farcs0022)\) at a position angle of \(171\arcdeg\), corresponding to \(890\pm10~\mathrm{pc}\) for the major axis.

\begin{figure}[ht!]
\epsscale{1.25}
\plotone{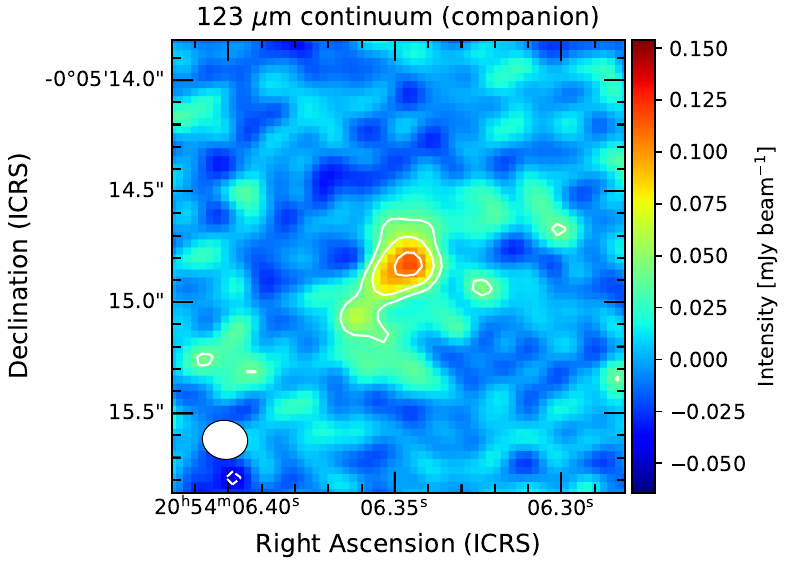}
\caption{Continuum image of the projected companion (the dashed rectangle in Figure \ref{fig:cont}). The contours are plotted at \((-3,3,5,8)\times\sigma\).\label{fig:comp}}
\end{figure}

We also found an additional continuum source positioned \(2\farcs4\) west of the quasar and detected at \(\mathrm{S/N}=8.9\) (Figure \ref{fig:comp}). Since OH is not detected there, it is not clear at this point whether the source is a physical companion or is at different redshift and happens to lie within the solid angle subtended by the primary beam. The projected separation from J2054-0005 corresponds to \(\approx14~\mathrm{kpc}\) if they are at the same redshift. The peak intensity is measured to be \(116\pm13~\mathrm{\mu Jy~beam^{-1}}\) at \((\alpha,\delta)_\mathrm{ICRS}=(20^\mathrm{h}54^\mathrm{m}06\fs344,-0\arcdeg05\arcmin14\farcs83)\), and the flux density is \(S_\nu(r<0\farcs5)=377\pm9~\mathrm{\mu Jy}\).

%%%%%
\subsection{OH gas}\label{subsec:ohf}

\begin{figure*}[ht!]
\epsscale{0.8}
\plotone{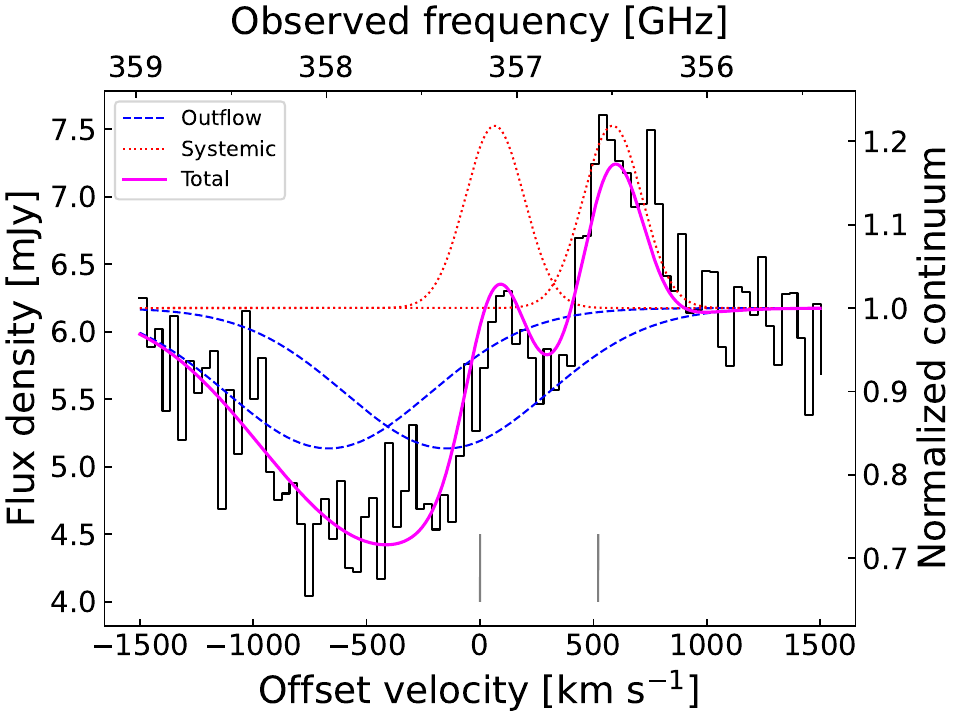}
\caption{Integrated OH \(119~\micron\) spectrum extracted from the region that includes all pixels where the \(123\mathrm{-}\micron\) continuum is detected at \(>3~\sigma\). The dashed blue and dotted red curves are the absorption and emission components, respectively, determined from two double-gaussian line fitting, and the solid magenta curve is their sum (see Section \ref{subsec:fit}). The offset velocity is measured with respect to the \(119.23~\micron\) line, as indicated by a vertical line.\label{fig:ohs}}
\end{figure*}

The OH \(119~\micron\) line is robustly detected toward the quasar. Figure \ref{fig:ohs} shows an integrated OH spectrum (total flux density) extracted from the region where the \(123\mathrm{-}\micron\) (LSB) continuum is detected at \(>3~\sigma\). We selected this broad region because there is a possibility that OH is distributed throughout the galactic disk traced by dust. The OH profile is dominated by a broad absorption feature at negative velocities, and emission at near-systemic velocities, exhibiting a typical P-Cygni profile, such as the one observed toward the local ULIRG Mrk 231 \citep{Fis10}. The line is very broad: what appears to be either OH absorption or emission extends over a continuous velocity from \(-1500~\mathrm{km~s^{-1}}\) to \(+1000~\mathrm{km~s^{-1}}\).

We can estimate the optical depth \(\tau\) as long as the line is not completely opaque. If emission at velocities of the absorption line is negligible, the observed flux density is \(S(v)=S_\mathrm{cont}e^{-\tau}\), where \(S_\mathrm{cont}\) is the continuum flux density, hence

\begin{equation}\label{eq:tau}
\tau(v)=-\ln\left[\frac{S(v)}{S_\mathrm{cont}}\right]
\end{equation}
Although there are reasons to assume that OH is not optically thin, e.g., if the gas distribution is clumpy and the absorbing gas does not cover the continuum entirely, the apparent optical depth at the line center, where absorption is maximum, is \(\tau\approx0.36\).

In the vicinity of the OH doublet, there are CH\(^{+}\) (\(J=3\mathrm{-}2\)) at the sky frequency of 355.364 GHz, and \(^{18}\)OH (\(J=5/2\mathrm{-}3/2\)) at 355.021 GHz, that may be responsible for the decrease in flux that appears at the offset velocity \(v\gtrsim1000\) km s\(^{-1}\). A similar feature attributed to the \(^{18}\)OH line is found in Mrk 231, indicating an enhanced \([^{18}\mathrm{OH}]/[\mathrm{OH}]\) abundance due to processing by star formation \citep{Fis10,GA14}. However, the lines are sufficiently separated from OH and unlikely to significantly affect the analysis of the line profile described below.

%%%%%
\subsection{OH line fitting}\label{subsec:fit}

To investigate the kinematics of OH gas, we performed a least-squares fitting of the line profile using Python (\texttt{scipy.optimize.curve\_fit}). Since the line is a doublet, and there may be multiple components (systemic, outflow, or inflow), fitting was conducted using two double-gaussian functions. Although the spectrum is likely to be more complicated than this simple structure, we aimed at limiting the number of free parameters while extracting key quantities related to the analysis of outflows. The fitting constraints were the following: the separation between the doublet lines of a double gaussian is fixed to \(521~\mathrm{km~s^{-1}}\), and their peak values and FWHMs are set to be equal (e.g., \citealt{Goi02}). The continuum intensity within the velocity range covered by the two adjacent USB spectral windows was assumed to be constant. On the other hand, the line intensity, continuum intensity, and the velocity separation of the double gaussians of different components (e.g., systemic and outflow) were set as free parameters.

The results of fitting are shown in Figure \ref{fig:ohs} and listed in Table \ref{tab:com}. We find OH emission, traced by a double gaussian with an FWHM line width of \(306\pm55~\mathrm{km~s^{-1}}\) near the systemic velocity, and a broad absorption feature with \(\mathrm{FWHM}=1052\pm234~\mathrm{km~s^{-1}}\) with the peak absorption velocity of \(v_\mathrm{cen}=-669\pm87~\mathrm{km~s^{-1}}\) relative to the systemic velocity. The FWHM of the emission line obtained from fitting is comparable to those of the emission lines of [\ion{C}{2}] 158 \(\micron\) (\(243\pm10~\mathrm{km~s^{-1}}\); \citealt{Wan13}), [\ion{O}{3}] 88 \(\micron\) (\(282\pm17~\mathrm{km~s}^{-1}\); \citealt{Has19}), and CO (\(J=6\rightarrow5\)) (\(360\pm110~\mathrm{km~s}^{-1}\); \citealt{Wan10}). However, [\ion{O}{3}] 88 \(\micron\) is a tracer of ionized gas, and it is unclear whether [\ion{C}{2}] 158 \(\micron\) is traces the same molecular medium as OH does. Only CO is directly comparable to OH as a molecular gas tracer, and the line widths of the two are in agreement with each other.

The terminal velocity (maximum extent of the blue-shifted wing of the absorption) is at least \(v_\mathrm{max}\approx-1500~\mathrm{km~s^{-1}}\) but may be beyond the spectral coverage. Another indicator of terminal velocity is \(v_{98}\), the velocity above which \(98\%\) of absorption takes place. This quantity is found to be \(v_{98}=-1574\pm35~\mathrm{km~s^{-1}}\). On the other hand, the velocity above which \(84\%\) of absorption takes place, is \(v_{84}=-1104\pm35~\mathrm{km~s^{-1}}\). The fitted emission components are red-shifted relative to the systemic by \(65\pm15~\mathrm{km~s}^{-1}\). This is not unusual, as such positive shifts in emission components have been observed in the majority of nearby galaxies that exhibit P-Cygni profiles and likely arise from outflows on the opposite side of the continuum (receding relative to the observer) \citep{Vei13}. The uncertainties above include only those from fitting. The redshift uncertainty expressed in velocity is \(\approx8~\mathrm{km~s}^{-1}\).

The continuum flux density in the USB spectral windows was found by fitting to be \(S_\mathrm{cont}=6.175\pm0.092~\mathrm{mJy}\). This is \(\approx8\%\) higher than the LSB continuum (see Section \ref{subsec:con}), but not unexpected, because the continuum emission at this frequency is in the Rayleigh-Jeans domain. Also, the regions where the fluxes were extracted (LSB continuum within \(r<0\farcs5\), USB continuum where LSB continuum is detected at \(>3~\sigma\)) are not equal, but are similar in size. Given the fact that there are almost no line-free channels in the USB spectrum and that fitting was done under simple assumptions, we find this to be a reasonable estimate.

\begin{figure}[ht!]
\epsscale{1.2}
\plotone{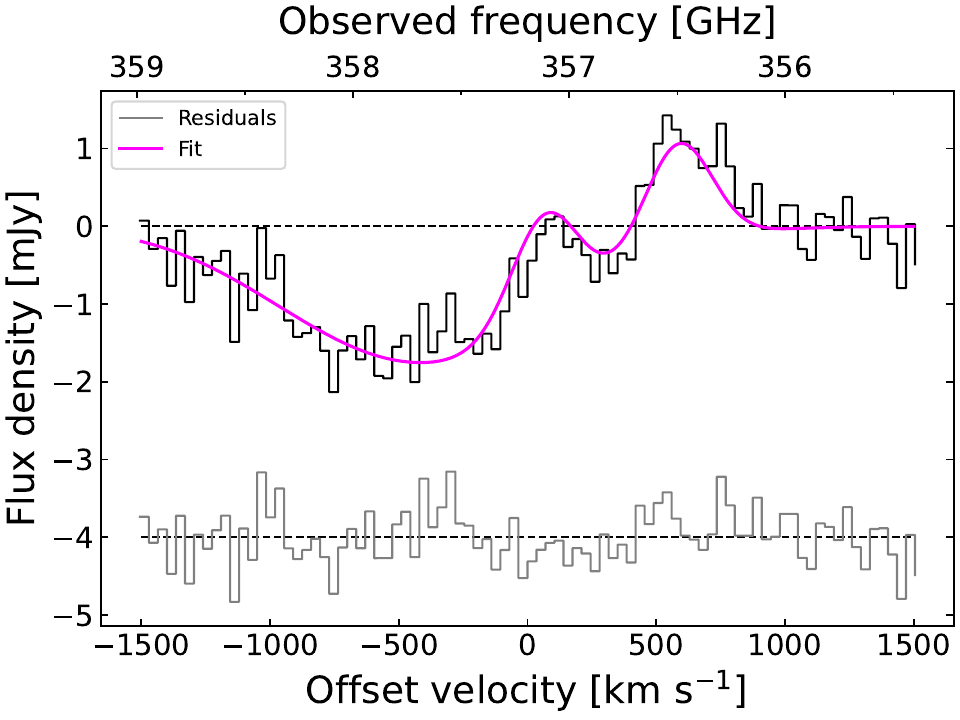}
\caption{Continuum-subtracted OH \(119~\micron\) spectrum with the computed fit (as in Figure \ref{fig:ohs}) and its residuals shown at the bottom.\label{fig:res}}
\end{figure}

Figure \ref{fig:res} shows a continuum-subtracted OH spectrum, including the best fit and fitting residuals.

\begin{deluxetable*}{lcc}[ht!]
\tablecaption{OH Line Fit Parameters\label{tab:com}}
\tablewidth{0pt}
\tablehead{
\colhead{} & \colhead{Absorption (outflow)} & \colhead{Emission (systemic)}
}
\startdata
Peak value \(S_\mathrm{max}\) [\(\mathrm{mJy}\)] & \(-1.039\pm0.078\) & \(1.35\pm0.27\)  \\
Integrated flux \(\mathcal{S}_\mathrm{OH}\) [\(\mathrm{Jy~km~s^{-1}}\)] & \(-1.16\pm0.27\) & \(0.44\pm0.12\) \\
Center velocity \(v_\mathrm{cen}\) [\(\mathrm{km~s^{-1}}\)] & \(-669\pm87\) & \(65\pm15\) \\
\(v_{84}~[\mathrm{km~s^{-1}}]\) & \(-1104\pm35\) & ... \\
\(v_{98}~[\mathrm{km~s^{-1}}]\) & \(-1574\pm35\) & ... \\
Standard deviation \(\sigma_v\) [\(\mathrm{km~s^{-1}}\)] & \(446\pm99\) & \(129\pm23\) \\
FWHM line width [\(\mathrm{km~s^{-1}}\)] & \(1052\pm234\) & \(306\pm55\) \\
Equivalent width \(W~[\mathrm{km~s^{-1}}]\) & \(200\pm44\) & \(-66\pm19\) \\
\enddata
\tablecomments{All quantities are calculated from gaussian fits. Here, \(\mathrm{FWHM}=\sqrt{8\ln{2}}\sigma_v\), \(\mathcal{S}_\mathrm{OH}=\sqrt{2\pi}\sigma_vS_\mathrm{max}\), and \(W\) is calculated according to Equation (\ref{eq:eqw}). The quantities \(\mathcal{S}_\mathrm{OH}\) and \(W\) are given for a single line of the doublet (the total equivalent width is twice the value). The continuum flux density is \(S_\mathrm{cont}=6.175\pm0.092~\mathrm{mJy}\).}
\end{deluxetable*}

%%%%%%%%%%%%%%%%%%%%%%%
\section{Molecular gas outflow}\label{sec:out}
%%%%%%%%%%%%%%%%%%%%%%%

The absorption line is tracing the OH gas between the continuum source and the observer. Since this is observed in blueshift relative to the host galaxy, the line reveals unambiguous signature of an outflow, as it has been observed toward AGNs and star-forming galaxies at lower \(z\) (e.g., \citealt{Vei13}). For example, the nearby ULIRG Mrk 231 exhibits a similar terminal velocity of \(\approx-1500~\mathrm{km~s^{-1}}\) and a P-Cygni line profile, which are attributed to an AGN-driven outflow \citep{Fis10,Spo13}. Since we do not detect red-shifted absorption, there is no significant molecular gas inflow in J2054-0005 along the line of sight. In the analysis below, we consider the absorption to be tracing an outflow, and derive the basic properties, such as mass, mass outflow rate, and kinetic energy.

\subsection{Outflow mass}\label{subsec:mas}

Assuming that the outflow has the shape of a thin spherical shell of radius \(r_\mathrm{out}\), the molecular gas mass in the outflow (including helium and heavier elements) can be expressed as \citep{Vei20}

\begin{equation}\label{eq:ofm}
M_\mathrm{out}=\mu m_\mathrm{H}N_\mathrm{H}\Omega r_\mathrm{out}^2,
\end{equation}
where \(\mu=1.36\) is the mean particle mass per hydrogen nucleus, \(m_\mathrm{H}\) is the hydrogen atom mass (in kilograms), \(N_\mathrm{H}=2N_\mathrm{H_2}\) is the column density of hydrogen nuclei (in \(\mathrm{m}^{-2}\)), and \(\Omega\) is the solid angle (``opening angle'') subtended by the shell \citep{Vei20}. Thus, the term \(\Omega r_\mathrm{out}^2\) is the area of the outflowing shell, and the opening angle is given by  \(\Omega=4\pi f\), where \(f\) is the dimensionless covering factor, i.e., the fraction of the sphere covered by the outflow as seen from its origin.

The column density \(N_\mathrm{H_2}\) cannot be measured directly, but we can find the column density of OH molecules (\(N_\mathrm{OH}\)) and then apply a \([\mathrm{OH}]/[\mathrm{H_2}]\) abundance ratio to get \(N_\mathrm{H_2}\). \(N_\mathrm{OH}\) can be calculated from the measured absorption line under the approximation of local thermodynamic equilibrium (LTE). The OH column density can be expressed as (e.g., \citealt{MS15})

\begin{equation}\label{eq:cmd}
N_\mathrm{OH}=\frac{8\pi}{\lambda_{ul}^3A_{ul}}\frac{Q}{g_u}\frac{e^{E_l/kT_\mathrm{ex}}}{1-e^{-\Delta E/kT_\mathrm{ex}}}\int\tau(v) dv,
\end{equation}
where \(\lambda_{ul}=119.23~\micron\) is the rest-frame wavelength, \(A_{ul}=0.1388~\mathrm{s}^{-1}\) is the Einstein coefficient for the \(u\rightarrow l\) transition \(J=5/2\rightarrow3/2\) \citep{Sch05}, \(g_u=6\) is the statistical weight of the \(J=5/2\) level, \(\Delta E=E_u-E_l\) is the energy difference of the two levels (\(E_l=0\)), \(T_\mathrm{ex}\) is the excitation temperature, and \(Q\) is the partition function. If \(\tau\ll1\), the integral is approximately equal to the equivalent width, defined as \(W=\int(1-e^{-\tau})dv\). Assuming \(T_\mathrm{ex}=50~\mathrm{K}\), which is equal to the dust temperature (\(T_\mathrm{d}=50\pm2~\mathrm{K}\); \citealt{Has19}), we get \(Q\simeq19\) \citep{Pic98}.  The term \(Q(1-e^{-\Delta E/kT_\mathrm{ex}})^{-1}\) increases by a factor of \(\approx3\) from 50 to 150 K. Note that the equivalent width in Equation (\ref{eq:cmd}) is calculated for a single line of the doublet (\(W\) in Table \ref{tab:com}) because \(Q\) accounts for the \(\Lambda\)-doubling.

The OH abundance has been measured for the Milky Way galaxy (e.g., \citealt{Goi02,Goi06,Ngu18,Rug18}) and a number of nearby galaxies where multiple OH lines were detected \citep{Spi05,Fal15,Sto18}, and it has been analyzed in theoretical work employing simulations \citep{RFG18}. A relatively high value of \([\mathrm{OH}]/[\mathrm{H_2}]=5\times10^{-6}\) is reported for Sgr B2 \citep{Goi02}. Although this value is often used to derive H\(_2\) mass from OH, it may be lower at sub-solar metallicities that may be applicable to high-\(z\) sources. This is, however, not necessarily the case for evolved systems, as near-solar metallicities have been found in some quasars at \(z>6\) \citep{Nov19,Li20b,Ono20}. On the other hand, a low abundance of \([\mathrm{OH}]/[\mathrm{H_2}]\approx1\times10^{-7}\) has been reported recently even for various regions in the Galaxy \citep{Ngu18,Rug18}. We derive the outflow properties using \([\mathrm{OH}]/[\mathrm{H_2}]=5\times10^{-7}\) (Table \ref{tab:out}) as a reasonable compromise between the two extremes. Choosing this value is also motivated by the fact that it yields an outflow mass and mass outflow rate comparable to those obtained using an empirical relation presented in Section \ref{subsec:emp} below.

The radius of the outflow (\(r_\mathrm{out}\)), as given in Equation \ref{eq:ofm}, is the radius of a thin outflow shell \citep{Vei20}. We adopt a radius of \(r_\mathrm{out}=2~\mathrm{kpc}\), which is the size of the \(123~\micron\) continuum shown in Figure \ref{fig:cont} and the region from which the spectrum in Figure \ref{fig:ohs} is extracted. This radius is consistent with discussion in Section \ref{subsec:res} below, where it is argued that the OH outflow is extended to at least \(r\gtrsim1~\mathrm{kpc}\) from the center.

For the covering factor, we use \(f=0.3\) in the analysis below. This value is also adopted in \citet{Spi20b} for a sample of star-forming galaxies at redshift \(z=4\mathrm{-}5\), although it can be as large as \(f\approx0.8\) \citep{Spi18}. Even if we adopt \(f=1\), the main results of the scaling relations discussed in Section \ref{sec:dis} do not change, albeit the mass outflow rates would be larger by a factor of \(\approx3\).

The integral in Equation (\ref{eq:cmd}) is calculated by inserting \(\tau\) from Equation (\ref{eq:tau}),

\begin{equation}\label{eq:eqw}
W=-\int\ln\left[\frac{S(v)}{S_\mathrm{cont}}\right]dv,
\end{equation}
where \(S(v)\) is the profile obtained from gaussian fitting and \(S_\mathrm{cont}\) is a constant (Table \ref{tab:com}). The equivalent width of a single line of the doublet is \(W=200~\mathrm{km~s^{-1}}\), yielding \(M_\mathrm{out}\approx4.9\times10^9~M_\odot\). The obtained \(W\) is larger than that found in almost all nearby ULIRGs \citep{Vei13} and dusty star-forming galaxies at \(z=4\mathrm{-}5\) \citep{Spi20a}, and is largest in a quasar at \(z>6\) reported to date. The calculated outflow gas mass and other dynamical quantities (derived below) are listed in Table \ref{tab:out}. The outflow mass is \(\sim8\mathrm{-}16\%\) of the total molecular gas mass \citep{Dec22}.

\begin{deluxetable*}{lccc}[ht!]
\tablecaption{Molecular Outflow Properties\label{tab:out}}
\tablewidth{0pt}
\tablehead{
\colhead{Quantity} & \colhead{LTE} & \colhead{Empirical relation}
}
\startdata
Column density \(N_\mathrm{OH}~[\mathrm{cm^{-2}}]\) &  \(7.5\times10^{15}\) & ... \\
OH abundance \([\mathrm{OH}]/[\mathrm{H}_2]\) & \(5\times10^{-7}\) & ... \\
Column density \(N_\mathrm{H_2}~[\mathrm{cm^{-2}}]\) & \(1.5\times10^{22}\) & ... \\
Mass \(M_\mathrm{out}\) [\(M_\odot\)] & \(4.9\times10^9\) & \(4.5\times10^9\) \\
Mass outflow rate \(\dot{M}_\mathrm{out}\) [\(M_\odot~\mathrm{yr^{-1}}\)] & \(1700\) & \(1500\) \\
Mass loading factor \(\eta\) & \(2.2\) & \(1.9\) \\
Depletion time \(t_\mathrm{dep}\) [yr] & \((2\mathrm{-}4)\times10^7\) & \((2\mathrm{-}4)\times10^7\) \\
Kinetic energy \(E_\mathrm{out}~[\mathrm{erg}]\) & \(2.2\times10^{58}\) & \(2.0\times10^{58}\) \\
Power \(\dot{E}_\mathrm{out}~[L_\odot]\) & \(6.2\times10^{10}\) & \(5.7\times10^{10}\) \\
\enddata
\tablecomments{The LTE values are calculated using the covering factor \(f=0.3\), outflow radius \(r_\mathrm{out}=2~\mathrm{kpc}\), excitation temperature \(T_\mathrm{ex}=50~\mathrm{K}\), and outflow velocity \(v_\mathrm{out}=669~\mathrm{km~s^{-1}}\). The empirical relation for the mass outflow rate is given in Equation (\ref{eq:emp}).}
\end{deluxetable*}

%%%%%
\subsection{Mass outflow rate}\label{subsec:mor}

\subsubsection{Optically-thin outflow model}\label{subsec:oto}

Assuming that the outflow is expanding at velocity \(v_\mathrm{out}\) as a thin spherical shell, the mass outflow rate averaged over the outflow lifetime can be expressed as

\begin{equation}\label{eq:mor}
\dot{M}_\mathrm{out}=M_\mathrm{out}\frac{v_\mathrm{out}}{r_\mathrm{out}}.
\end{equation}
This equation gives a conservative estimate \citep{Mai12,Lut20,Sal20,Vei20}. For the outflow velocity, we adopt \(v_\mathrm{out}=669~\mathrm{km~s^{-1}}\), based on the center velocity (\(v_\mathrm{cen}\)) of the absorption feature (see Table \ref{tab:com} and Section \ref{subsec:fit}). Although this is the mean value along the line of sight, it is equal to the outflow velocity in the case of a spherically-symmetric outflow. Taking \(W=200~\mathrm{km~s^{-1}}\) (equivalent width of the absorption line), \([\mathrm{OH}]/[\mathrm{H_2}]=5\times10^{-6}\), \(r_\mathrm{out}=2~\mathrm{kpc}\), and \(f=0.3\), we obtain a lower limit of \(\dot{M}_\mathrm{out}>168~M_\odot~\mathrm{yr}^{-1}\). Note that if we adopt a low abundance of \([\mathrm{OH}]/[\mathrm{H_2}]=1\times10^{-7}\) and \(f=1\), the upper limit of the mass outflow rate becomes \(\dot{M}_\mathrm{out}\approx2.8\times10^4~M_\sun~\mathrm{yr}^{-1}\). However, the upper limit yields an outflow mass that exceeds the total molecular gas mass in the host galaxy \citep{Dec22}. Clearly, the uncertainty of the mass outflow rate is dominated by the poorly constrained OH abundance. We adopt a moderate abundance of \([\mathrm{OH}]/[\mathrm{H_2}]=5\times10^{-7}\) and \(f=0.3\) in the analysis below.

The dynamical age of the outflow is defined as the time needed for outflowing gas to travel a distance \(r_\mathrm{out}\) at a constant velocity \(v_\mathrm{out}\). Using the derived quantities above, the outflow age is \(t_\mathrm{out}=r_\mathrm{out}/v_\mathrm{out}\sim3\times10^6~\mathrm{yr}\). The timescale is much shorter than the depletion time due to star formation (\(\sim4\mathrm{-}8\times10^7~\mathrm{yr}\)).

\subsubsection{Empirical relation}\label{subsec:emp}

Alternatively, we can circumvent the above assumptions and use an empirical formula for the mass outflow rate discussed in the literature, hoping that it is applicable to the EoR quasar. The ``recipe'' formula from \citet{HC20} modified by \citet{Spi20b} takes the form

\begin{equation}\label{eq:emp}
\left(\frac{\dot{M}^\mathrm{emp}_\mathrm{out}}{M_\odot~\mathrm{yr}^{-1}}\right)=1.4\left(\frac{W_{v<-200}}{\mathrm{km~s^{-1}}}\right)\left(\frac{L_\mathrm{IR}}{10^{12}L_\sun}\right)^{1/2}+180,
\end{equation}
where \(W_{v<-200}\) is the OH \(119~\micron\) equivalent width at \(v<-200~\mathrm{km~s^{-1}}\). This equivalent width, derived from the observed spectrum, is \(W_{v<-200}\approx268~\mathrm{km~s^{-1}}\), yielding a mass outflow rate of \(\dot{M}^\mathrm{emp}_\mathrm{out}\approx1500~M_\sun\). The value is significantly larger compared to those for the star-forming galaxies at redshift \(z=4\mathrm{-}5\) reported in \citet{Spi20b}, which have values between \(220\) and \(1290~M_\odot~\mathrm{yr}^{-1}\).

Using the mass outflow rate from Equation (\ref{eq:emp}), we calculate the outflow mass \(M_\mathrm{out}^\mathrm{emp}=(r_\mathrm{out}/v_\mathrm{out})\dot{M}_\mathrm{out}^\mathrm{emp}\) and other dynamical quantities. All outflow properties derived from this empirical relation are listed in Table \ref{tab:out}.

Some caveats of this approach include the fact that Equation (\ref{eq:emp}) only incorporates the equivalent width at \(v<-200~\mathrm{km~s^{-1}}\) (to exclude the systemic component assuming that it does not exceed this velocity) regardless of the line width of the outflow-tracing absorption line and the mean outflow velocity. None the less, the mass outflow rate and outflow mass obtained using Equation (\ref{eq:emp}) are in agreement with those obtained under LTE and moderate OH abundance  (Table \ref{tab:out}).

%%%%%
\subsection{Resolved OH absorption and emission}\label{subsec:res}

\begin{figure}[ht!]
\epsscale{1.2}
\plotone{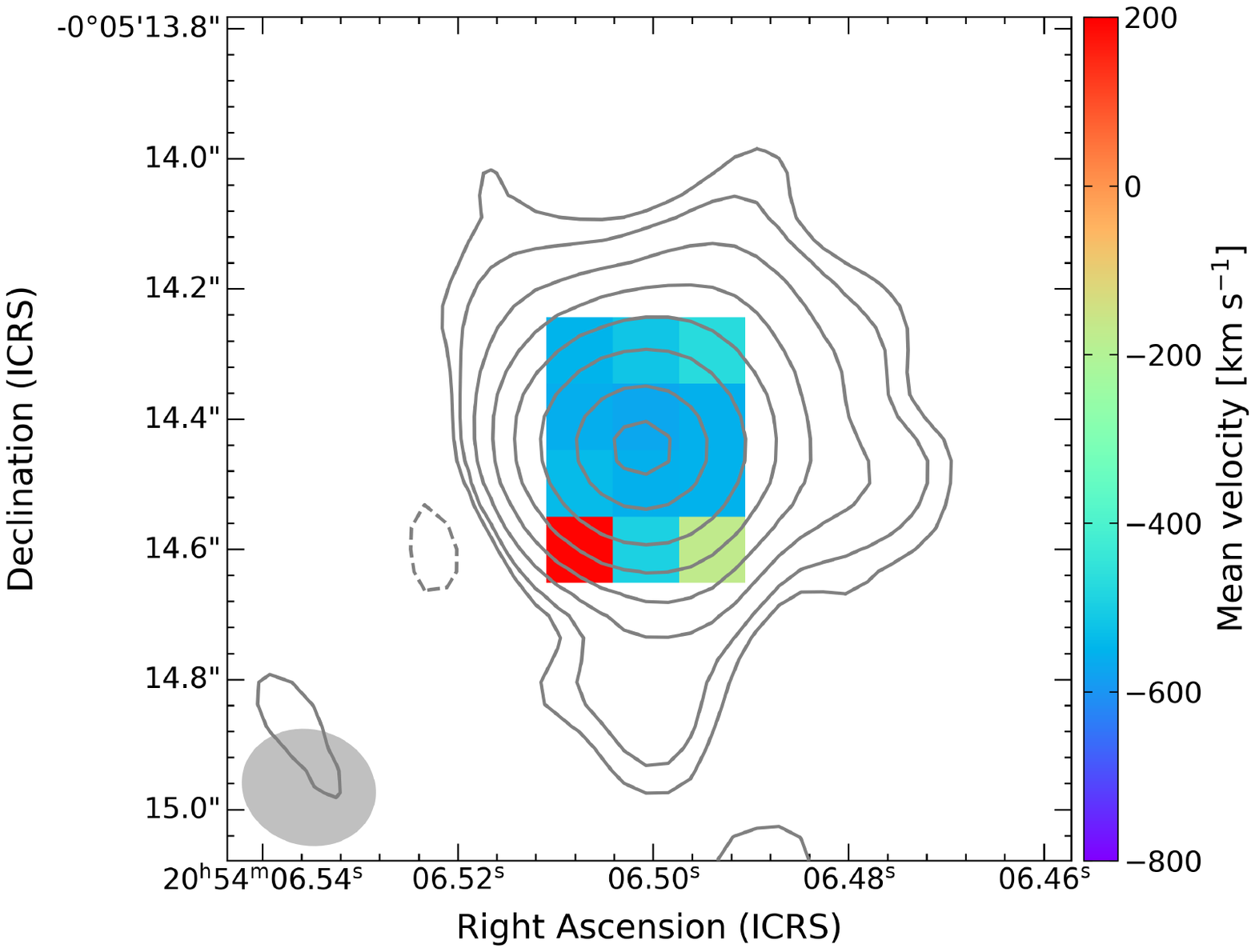}
\caption{Moment 1 image of the OH data calculated within the velocity range of \([-1505,420]~\mathrm{km~s}^{-1}\) that exhibits OH absorption. The spectra from each pixel are shown in Figure \ref{fig:spe}. The contours are the continuum as in Figure \ref{fig:cont}.\label{fig:mom}}
\end{figure}

\begin{figure*}[ht!]
\epsscale{1.2}
\plotone{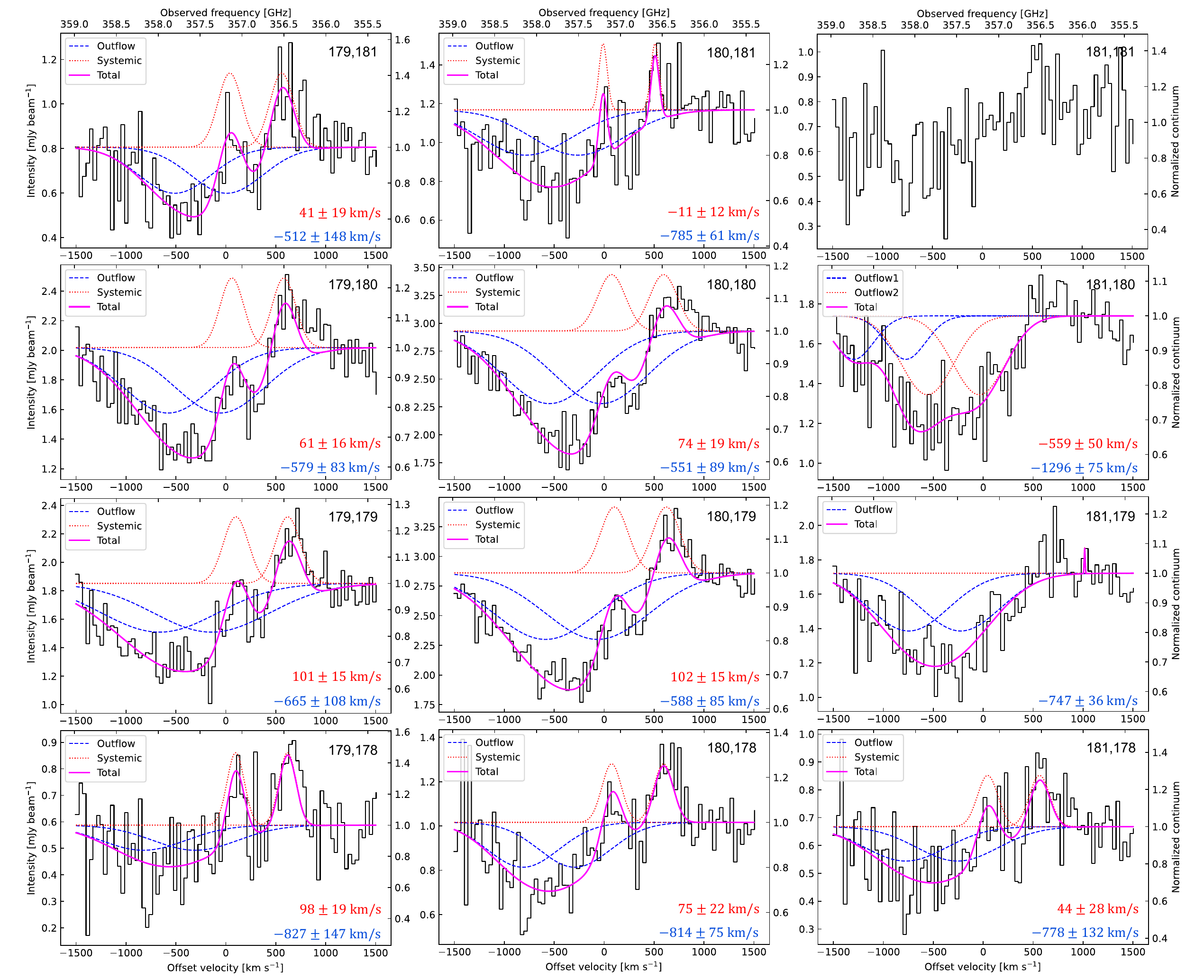}
\caption{Fitted OH \(119~\micron\) spectra (profile map) toward the central \(\approx1.7\times2.3~\mathrm{kpc^2}\) region extracted from the pixels in Figure \ref{fig:mom}. Each pixel has the area \(0\farcs1\times0\farcs1\) (approximately one half of the beam size) with offset \(0\farcs1\). The spectra are denoted by the pixel coordinates \((x,y)\) at the top right. The continuum peak is approximately between (180,179) and (180,180). The center velocities of the fitted emission and absorption lines are shown at the bottom right of each spectrum. The narrow emission feature at (181,179) is likely an artifact.\label{fig:spe}}
\end{figure*}

The high angular resolution (\(\approx1~\mathrm{kpc}\)) and sensitivity allow us to probe the spatial distribution of OH gas velocity for the first time in a quasar at \(z>6\). We extracted OH spectra from adjacent rectangular regions of area \(0\farcs1\times0\farcs1\), corresponding to approximately one half of the synthesized beam, and performed double-gaussian fitting in each region using the procedure described in Section \ref{subsec:fit}. To obtain successful fits with a minimum number of free parameters, we fit all regions with only two double-gaussians. A moment 1 image of OH that shows the positions of the regions as pixels is shown in Figure \ref{fig:mom}. The spectra in Figure \ref{fig:spe} were extracted from the 12 pixels in Figure \ref{fig:mom}, labelled (179,178) in the bottom left corner, (181, 181) in the top right corner, etc., and shown as a \(3\times4\) profile map. All spectra that could yield reasonable fits are plotted in Figure \ref{fig:spe} together with the best fits. The OH doublet absorption was successfully fitted in 11 rectangular regions (Figure \ref{fig:spe} and Table \ref{tab:spe}). The profile could not be well-fitted in the surrounding regions, where the continuum intensity is weaker and OH is not significantly detected.

The absorption line is marginally spatially resolved, which can be inferred from a north-south shift in the peak absorption and emission velocities obtained by fitting. This implies that the distribution of OH gas is not confined to the AGN, which is too compact to be resolved, but extends over a broader (\(r\gtrsim1~\mathrm{kpc}\)) central region. The FWHM line widths (Table \ref{tab:spe}) are relatively comparable throughout the region (\(\approx800\mathrm{-}1300~\mathrm{km~s}^{-1}\)). The peak velocity is not minimum (most negative) at the continuum center, as may be expected from a spherically symmetric outflow emerging from the center, but retains comparable values throughout the map. Although the fitting uncertainties are large, the fitted absorption peak velocities appear to be more negative on the south side compared to the north side, though region (180,181) seems to deviate from this trend. The mean absorption velocities in each row in Figure \ref{fig:spe} from north to south are \((-648\pm80, -565\pm61, -667\pm47, -806\pm70)~\mathrm{km~s}^{-1}\), excluding (181,180). These results suggest that the outflow is not uniform and might have the shape of a cone whose axis is inclined with respect to the line of sight. Observations at higher resolution are needed to get a clearer picture of the outflow geometry.

OH is detected in emission at \(>3~\sigma\) in some regions and exhibits relatively comparable FWHM line widths throughout the region (\(\approx200\mathrm{-}330~\mathrm{km~s}^{-1}\)), consistent with reported [\ion{C}{2}] 158 \micron, \(\micron\) and [\ion{O}{3}] 88 \(\micron\), and CO (\(J=6\rightarrow5\)) line widths \citep{Wan10,Wan13,Has19}. This suggests that highly excited (warm or dense, shocked) molecular gas is distributed in the central 2 kpc region, either in the host galaxy or in the outflowing gas (the size of the region where emission could be fitted is \(\approx2~\mathrm{kpc}\) in diameter). The positive peak velocities of the emission lines are generally lower in the north compared to the south. The exception is at (181,178), though the line is only marginally detected there. The mean emission velocities in each row in Figure \ref{fig:spe} from north to south are \((15\pm11, 67\pm12, 101\pm11, 72\pm13)~\mathrm{km~s}^{-1}\).

Pixel (181,179) was fitted with an absorption feature and a very narrow emission feature (Figure \ref{fig:spe}). The latter is likely an artifact, because the line width is too narrow and the emission line is not significantly detected here. Pixel (181,180) was successfully fitted with two different absorption features. Since we did not put constraints on whether the fit should yield absorption or emission, the fitting turned out to be more successful with two absorption features than one absorption and one emission features at this pixel.

In order to investigate the origin of the apparent velocity shift in the OH emission line, we compare the OH data with [\ion{C}{2}] 158 \(\micron\) data. The moment 1 image in \citet{Wan13} shows that the [\ion{C}{2}] 158 \(\micron\) line exhibits a velocity gradient in the northwest-southeast direction, that is generally consistent with the north-south velocity increase in the fitted OH emission line spectra, albeit with an offset: the fitted OH emission is systematically redder than [\ion{C}{2}] 158 \(\micron\). Thus, it is possible that the OH emission follows the velocity field of the bulk gas in the host galaxy traced by [\ion{C}{2}] 158 \(\micron\).

\begin{figure}[ht!]
\epsscale{1.2}
\plotone{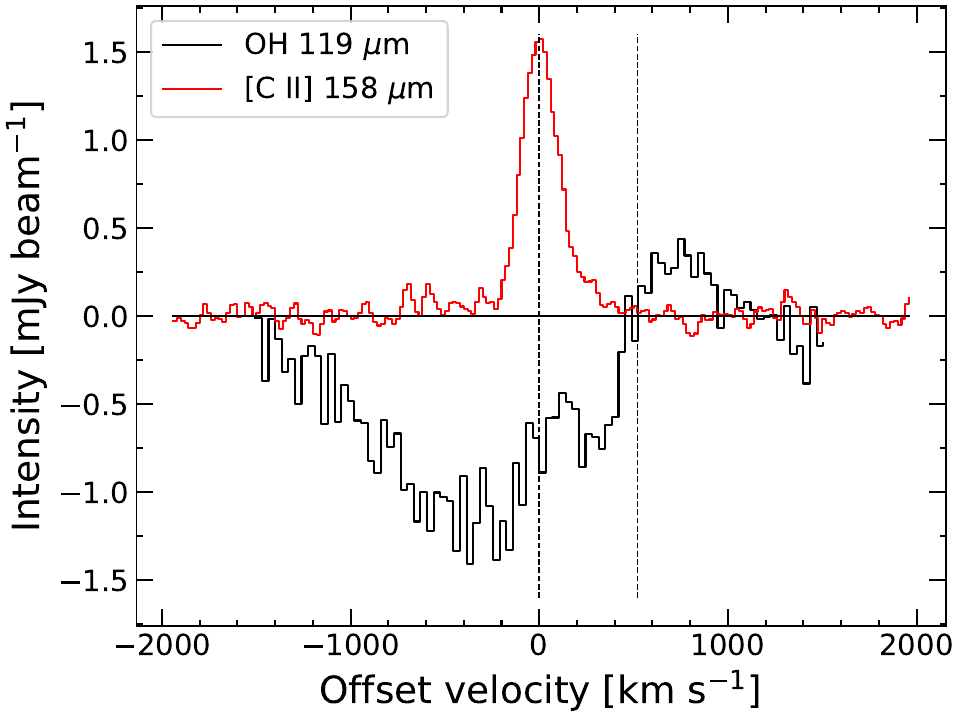}
\caption{Spectra of OH 119 \(\micron\) and [\ion{C}{2}] 158 \(\micron\) extracted from the central \(0.034\arcsec\) pixel (peak intensity of \(123~\micron\) continuum). The [\ion{C}{2}] intensity is scaled by \(\times0.2\). The vertical dashed lines show the OH doublet velocities.\label{fig:cii}}
\end{figure}

Figure \ref{fig:cii} shows a comparison of the OH \(119~\micron\) and [\ion{C}{2}] 158 \(\micron\) spectra (the [\ion{C}{2}] data are from \#2019.1.00672; S. Fujimoto, in prep.), extracted from the central pixel (maximum \(123~\micron\) continuum intensity; pixel size \(0\farcs034\)). These are the highest-resolution [\ion{C}{2}] data of this source and therefore best for comparison. In addition to the main emission profile, the [\ion{C}{2}] line profile exhibits a secondary component on the red-shifted side (up to \(+500~\mathrm{km~s}^{-1}\)), and there appears to be a blue-shifted component at velocities comparable to the OH outflow velocity (\(-600~\mathrm{km~s}^{-1}\)), although it is tentative at this choice of aperture. This is consistent with recent findings that OH \(119~\micron\) is a more robust tracer of outflows compared to [\ion{C}{2}] 158 \(\micron\) \citep{Spi20a}.

More details of the [\ion{C}{2}] observations and results will be presented in Fujimoto et al. (in prep.).

\begin{deluxetable*}{lcccc}[ht!]
\tablecaption{Fitting Results for Resolved Emission and Absorption Components\label{tab:spe}}
\tablewidth{0pt}
\tablehead{
\colhead{Region} & \colhead{\(v_\mathrm{cen}^\mathrm{emi}\) [\(\mathrm{km~s}^{-1}\)]} & \colhead{\(v_\mathrm{cen}^\mathrm{abs}\) [\(\mathrm{km~s}^{-1}\)]} & \colhead{FWHM\(^\mathrm{emi}\) [\(\mathrm{km~s}^{-1}\)]} & \colhead{FWHM\(^\mathrm{abs}\) [\(\mathrm{km~s}^{-1}\)]}
}
\startdata
179,178 & \(98\pm19\) & \(-827\pm147\) & \(202\pm56\) & \(1000\pm557\) \\
179,179 & \(101\pm15\) & \(-665\pm108\) & \(295\pm62\) & \(1368\pm338\) \\
179,180 & \(61\pm16\) & \(-579\pm83\) & \(278\pm67\) & \(1057\pm212\) \\
179,181 & \(41\pm19\) & \(-512\pm148\) & \(282\pm80\) & \(849\pm350\) \\
180,178 & \(75\pm22\) & \(-814\pm75\) & \(230\pm64\) & \(842\pm269\) \\
180,179 & \(102\pm15\) & \(-588\pm85\) & \(327\pm60\) & \(1224\pm223\) \\
180,180 & \(74\pm19\) & \(-551\pm89\) & \(330\pm82\) & \(1095\pm206\) \\
180,181 & \(-11\pm12\) & \(-785\pm61\) & \(96\pm31\) & \(1084\pm243\) \\
181,178 & \(44\pm28\) & \(-778\pm132\) & \(240\pm93\) & \(963\pm436\) \\
181,179 & ... & \(-747\pm36\) & ... & \(933\pm155\) \\
181,180 & ... & \(-559\pm50\), \(-1296\pm75\) & ... & \(621\pm125\), \(468\pm134\) \\
\enddata
\tablecomments{Regions are designated by image pixel numbers (\(x,y\)). Each pixel has area \(0\farcs1\times0\farcs1\), which is approximately one half of the synthesized beam. The spectrum at (181,180) could not be fitted with emission.}
\end{deluxetable*}

%%%%%%%%%%%%%%%%%
\section{Discussion}\label{sec:dis}
%%%%%%%%%%%%%%%%%

In this section, we first estimate the fractional contribution of AGN to the IR luminosity, and then discuss on the driving mechanism of the outflow, the fate of the outflowing gas, and its impact on the host galaxy.

%%%%%
\subsection{Fractional contribution of AGN to IR luminosity}\label{subsec:agn}

Estimating the SFR in high-\(z\) sources is often done by applying a conversion factor to the far-IR luminosity. However, in quasars, the IR luminosity is produced not just by dust heated by star formation, but also by dust heated by the AGN. To obtain a reliable estimate of SFR, it is therefore important to subtract the fractional contribution of the AGN (e.g., \citealt{Dur17}). Here, we applied a multi-wavelength spectral energy distribution (SED) modeling of the spectrum of J2054-0005 using the latest version of Code Investigating Galaxy Emission (\texttt{CIGALE}; \citealt{Boq19}). \texttt{CIGALE} is a state-of-the-art modeling and fitting code that combines a broad range of components, including a stellar population, AGN, and dust. For each parameter, \texttt{CIGALE} conducts a probability distribution function (PDF) analysis, yielding the output value as the likelihood-weighted mean of the PDF. For the SED fitting of J2054-0005, we used the following data: SDSS-z, WISE1, \emph{Herschel} (PACS, SPIRE), ALMA bands 6 and 7, and the upper limits from \emph{Herschel} \(500~\micron\) and the Very Large Array \(1.4~\mathrm{GHz}\) \citep{Ban15,Sha19}.

We assigned a flexible star formation history composed of a delayed component to account for a recent burst (e.g., \citealt{Don20}). The assumed values for the main stellar population are set to be between 300 and 750 Myr, with an \(e\)-folding time of 90 Myr, and a \citet{Cha03} IMF. The gas-phase metallicity was fixed to be two times lower than the solar value. This is motivated by the fact that the fitting result was somewhat more successful (in terms of \(\chi^2\)) with this value compared to that based on solar metallicity (e.g., \citealt{Nov19}), although both yielded a similar \(L_\mathrm{IR}\). Dust attenuation was modeled using a modified law from \citet{CF00}, and dust emission was modeled based on \citet{DL07}. The model assumes a mixture of grains exposed to variable radiation fields. We fixed the dust emission slope to \(\beta=2\), and allowed a range of radiation field intensities (\(10<U_\mathrm{min}<50\)). This approach can account for a very intense central heating source.

The AGN module used to determine the fractional contribution of the AGN to the total IR luminosity is based on \citet{FFH06}. The model takes into account three components through radiative transfer: the primary source located in the torus, the scattered emission by dust, and the thermal dust emission. We fixed the optical depth to \(2\), following some prescriptions from the literature (e.g., \citealt{CEF17}). For the sake of computational efficiency, we modeled two inclination angles of the torus (\(30\arcdeg\) and \(70\arcdeg\); \citealt{MGG19}).

\begin{figure*}[ht!]
\epsscale{0.9}
\plotone{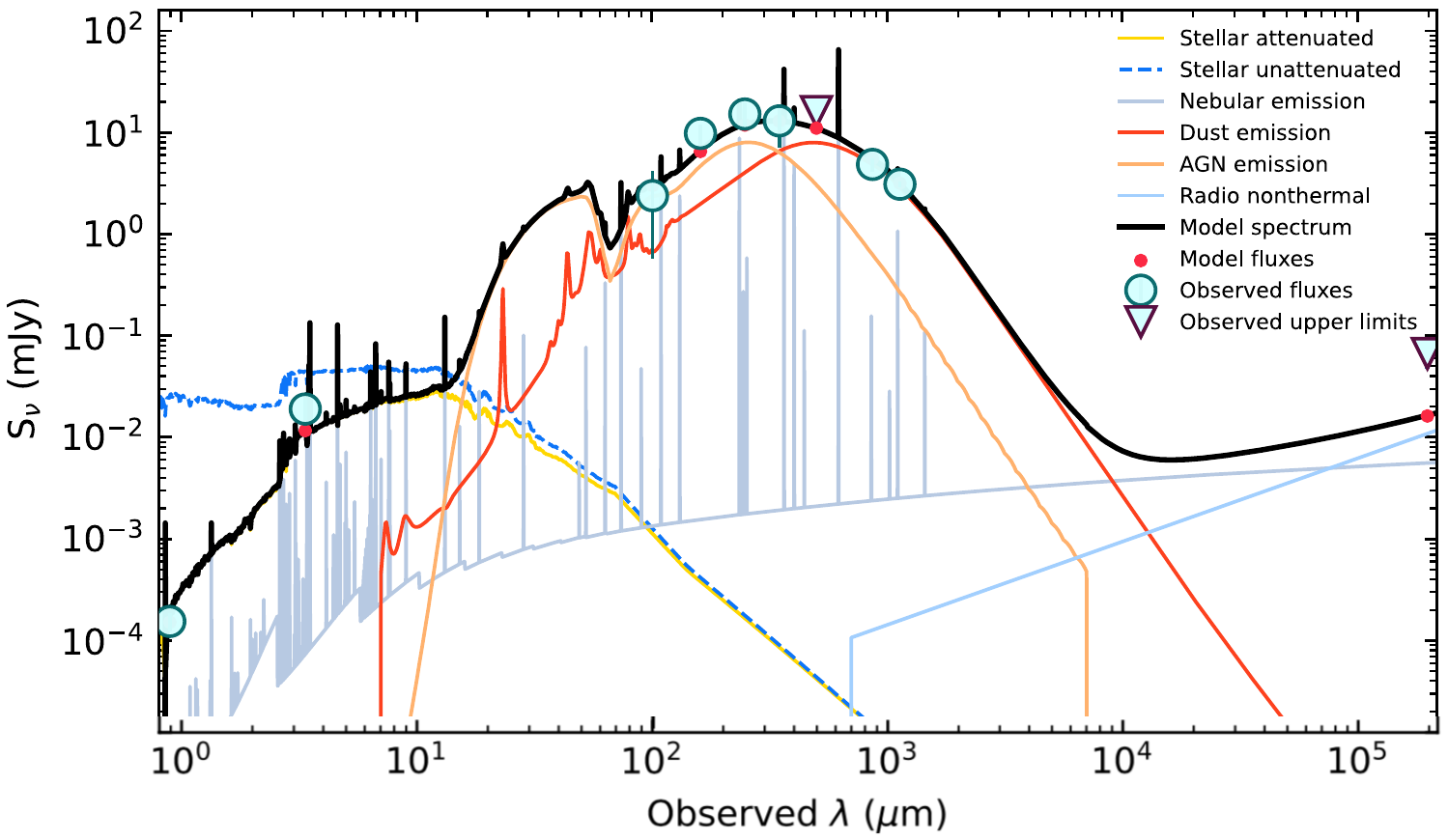}
\caption{Results of the SED fitting for J2054-0005 using the \texttt{CIGALE} code. The red line labeled ``Dust emission'' is the dust component heated by star formation, and the orange line labelled ``AGN emission'' is the contribution from the AGN. The latter makes up \(59\%\pm8\%\) of the total infrared luminosity. The fit has \(\chi^2=0.52\pm0.11\) (reduced). See the text for more details on the parameter setup.}\label{fig:agn}
\end{figure*}

The result of the procedure is shown in Figure \ref{fig:agn}. A reasonably-well fit was obtained, as indicated by a median reduced \(\chi^2=0.52\pm0.11\). We found that the total IR luminosity is \(L_\mathrm{IR}=(1.34\pm0.17)\times10^{13}~L_\sun\), consistent with previous studies \citep{Has19}, and that the fractional contribution of AGN to \(L_\mathrm{IR}\) is \(f_\mathrm{AGN}=0.59\pm0.08\). Thus, the AGN may be responsible for as much as \(\approx59\%\) of \(L_\mathrm{IR}\) in this source. This is comparable to the fractions reported for quasars at high \(z\) (e.g., \citealt{Sch15,Dur17,Tri22}). Accounting for this effect, the IR-derived star formation rate becomes \(\mathrm{SFR}=770\pm180~M_\odot~\mathrm{yr}^{-1}\), calculated using \(\mathrm{SFR}/M_\sun~\mathrm{yr}^{-1}=1.40\times10^{-10}L'_\mathrm{IR}/L_\sun\), where \(L'_\mathrm{IR}\) is the AGN-subtracted IR luminosity (integrated within \(\lambda_\mathrm{rest}=8-1000~\micron\)). The conversion factor corresponds to the Chabrier IMF, and was calculated by dividing the Kroupa IMF factor in \citet{Mur11} by \(1.06\) (e.g., \citealt{Zah12,Spe14}). We adopt this value in the analysis below.

The best fit was obtained for the torus inclination angle of \(30\arcdeg\). Based on low-\(z\) studies, this value is consistent with a Type 1 AGN (e.g., \citealt{Yan20}). It is also consistent with a relatively broad line width (\(\mathrm{FWHM}\approx4890~\mathrm{km~s}^{-1}\)) of Ly\(\alpha\) emission reported in \citet{Jia08}.

%%%%%
\subsection{Driving mechanism}\label{subsec:sva}

Is the outflow driven by star formation or does the AGN feedback from the accretion onto the supermassive black hole (SMBH) play a role? One way to address this problem is to investigate the energy and momentum of the outflow and compare them to the expected input from star formation.

Using the outflow mass and velocity derived in Section \ref{sec:out}, we calculate the bulk kinetic energy of the outflowing gas. Adopting the molecular gas mass derived under LTE (Table \ref{tab:out}), the kinetic energy is

\begin{equation}
E_\mathrm{out}=\frac{1}{2}M_\mathrm{out}v_\mathrm{out}^2\approx2.2\times10^{58}\mathrm{erg}.
\end{equation} 
and the power required to drive the molecular outflow (kinetic power) is

\begin{equation}
\dot{E}_\mathrm{out}=\frac{1}{2}\dot{M}_\mathrm{out}v_\mathrm{out}^2\approx2.4\times10^{44}~\mathrm{erg~s}^{-1}.
\end{equation}
This is equivalent to \(\approx6.2\times10^{10}~L_\sun\), which is \(\approx0.5\%\) of the total infrared luminosity of the source. If other ISM phases (atomic and ionized gas) are present in the outflow, the energy and power are larger.

The total momentum of the molecular outflow is

\begin{equation}
p_\mathrm{out}=M_\mathrm{out}v_\mathrm{out}\approx3.3\times10^{12}~M_\odot~\mathrm{km~s^{-1}}.
\end{equation}
By comparison, the momentum injection by a typical core-collapse supernova (SN; mass \(m_0\approx10~M_\odot\), velocity \(v_0\approx3000~\mathrm{km~s^{-1}}\)) is of the order of \(p_0\approx3\times10^4~M_\odot~\mathrm{km~s^{-1}}\), and the total momentum of an outflow driven by SN explosions is \(p_\mathrm{SN}\approx p_0R_\mathrm{SN}t_\mathrm{SN}\), where \(R_\mathrm{SN}\) is the SN rate, and \(t_\mathrm{SN}\) is the time interval measured from the onset of SN explosions. Thus, \(R_\mathrm{SN}t_\mathrm{SN}\) is the total number of SN explosions over the starburst episode. Adopting a relation between the SN rate and star formation rate, \(R_\mathrm{SN}/\mathrm{yr}^{-1}\approx\alpha_\mathrm{SN}(\mathrm{SFR}/{M_\odot~\mathrm{yr^{-1}}})\), where \(\alpha_\mathrm{SN}\approx0.01\mathrm{-}0.02\) depends on IMF and assumes continuous star formation \citep{Vei20}, and \(\mathrm{SFR}=770~M_\odot~\mathrm{yr^{-1}}\), we obtain \(R_\mathrm{SN}\approx15~\mathrm{yr^{-1}}\) (for \(\alpha=0.02\)). If the time interval of the SN feedback is equal to the dynamical age of the outflow (\(t_\mathrm{SN}= t_\mathrm{out}\)), the total SN momentum becomes \(p_\mathrm{SN}\approx1.3\times10^{12}~M_\odot~\mathrm{km~s^{-1}}\), produced by \(R_\mathrm{SN}t_\mathrm{SN}\approx4.5\times10^7\) SN explosions. Theoretical works suggest that the final momentum input per SN may be even higher (e.g., \citealt{KO15,WN15}). Moreover, the total momentum could be larger by a factor of two if stellar winds (radiation pressure) from massive stars play a significant role (e.g., \citealt{Lei99,MQT10}). Taking into account the possibility that other gas phases also participate in the outflow, so that the total momentum (molecular, neutral atomic, and ionized gas) is somewhat larger, it seems that star formation activity alone may be approximately sufficient to explain the observed outflow. Although we obtain this result based on a moderate relative OH abundance, a similar conclusion is reached if the empirical relation for the mass outflow rate is used.

On the other hand, the mean outflow velocity of \(670~\mathrm{km~s^{-1}}\) and the terminal velocity of \(1500~\mathrm{km~s^{-1}}\) exceed the outflow velocities typically measured in star-forming galaxies, which are found to be \(100\mathrm{-}500~\mathrm{km~s^{-1}}\) and \(<1000~\mathrm{km~s^{-1}}\), respectively (e.g., \citealt{Sug19}), although the outflow velocities are found to be higher in more extreme systems \citep{Spi20a}. By contrast, terminal velocities of outflows in AGN-dominated systems are often found to \(\gtrsim1000~\mathrm{km~s^{-1}}\) and as large as \(1500~\mathrm{km~s^{-1}}\) (e.g., \citealt{Rup05,Stu11,Spo13,Vei13,Gin20}). Theoretical studies also reproduce the velocities of \(\gtrsim1000~\mathrm{km~s}^{-1}\) and mass outflow rates of \(\sim10^3~M_\sun~\mathrm{yr}^{-1}\) in AGN-driven outflows (e.g., \citealt{IFM18,Cos18}). The results indicate that the AGN feedback (radiation pressure) may play a role in boosting the velocity in J2054-0005.

\begin{figure}[ht!]
\epsscale{1.2}
\plotone{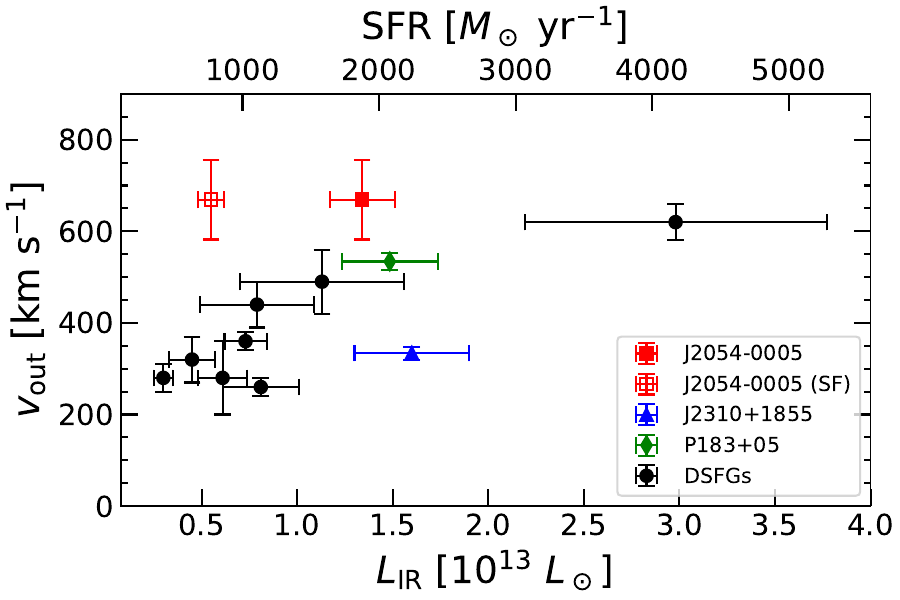}
\caption{Mean outflow velocity vs. total IR luminosity for quasars at \(z\approx6\) and dusty star-forming galaxies (DSFGs) at \(z=4\mathrm{-}5\), where OH outflows are detected \citep{Spi20a,But23}. The star formation rate is calculated as \(\mathrm{SFR}/M_\sun~\mathrm{yr}^{-1}=1.40\times10^{-10}L_\mathrm{IR}/L_\sun\), which is an upper limit for the quasars. The open square symbol shows J2054-0005 corrected for the AGN contribution. The data points for J2310+1855 and P183+05 are not AGN-corrected.\label{fig:v-LIR}}
\end{figure}

In Figure \ref{fig:v-LIR}, we show the mean line-of-sight outflow velocity plotted against the total IR luminosity (\(L_\mathrm{IR}\)) for 8 dusty star-forming galaxies (DSFGs) at \(z=4\mathrm{-}5\) and 3 quasars at \(z>6\). Here, \(L_\mathrm{IR}\) is obtained by integrating the flux over \(\lambda_\mathrm{rest}=8\mathrm{-}1000~\micron\) (Section \ref{subsec:agn}; \citealt{Sha19,Spi20a}), except P183+05, for which \(L_\mathrm{IR}=1.41L_\mathrm{FIR}\) \citep{Ven20}. For J2054-0005, we used \(L_\mathrm{IR}\) derived from the \texttt{CIGALE} fitting. The plot also shows the star formation rate, calculated using \(\mathrm{SFR}/M_\sun~\mathrm{yr}^{-1}=1.40\times10^{-10}L_\mathrm{IR}/L_\sun\), where a \citet{Cha03} IMF is assumed. The outflow in J2310+1855 was also detected in OH\(^{+}\) (\(1_1\leftarrow0_1\)) absorption \citep{Sha22} and the absorption velocity is comparable to that of OH plotted here.

Generally, there is an increase in \(v_\mathrm{out}\) with \(L_\mathrm{IR}\), though the relation is not clear for the quasars. One possibility is that this is simply because the AGN contribution in driving the outflow in J2054-0005 is relatively large, making the velocity higher than what it would be if star formation were the only driving mechanism. On the other hand, as discussed in \citet{But23}, if the outflows in these quasars are anisotropic (e.g., conical), it is possible that the random orientation results in no correlation because OH in most high-\(z\) sources is unresolved. Further observations including emission lines at higher resolution are necessary to clarify the outflow geometry.

%%%%%
\subsection{Suppression of star formation}\label{subsec:sfq}

The mass loading factor, defined as the ratio of the mass outflow rate to star formation rate, is an indicator of the outflow impact on star formation activity. In J2054-0005, using the AGN-corrected SFR calculated in Section \ref{subsec:agn}, we find

\begin{equation}
\eta=\frac{\dot{M}_\mathrm{out}}{\mathrm{SFR}}\sim2,
\end{equation}
implying efficient suppression of star formation. The total molecular gas mass in this quasar was estimated to be \(M_\mathrm{mol}=(3\mathrm{-}6)\times10^{10}M_\sun\) by \citet{Dec22} from a variety of tracers (dust, CO, [\ion{C}{1}] 609 \(\micron\), and [\ion{C}{2}] 158 \(\micron\) emission). Using these values, we calculate the depletion time, i.e., the time it takes for the outflow to remove molecular gas from the galactic center region, as

\begin{equation}
t_\mathrm{dep}=\frac{M_\mathrm{mol}}{\dot{M}_\mathrm{out}}\approx(2\mathrm{-}4)\times10^7~\mathrm{yr}.
\end{equation}

The relatively short timescale, as compared to the time required for star formation to consume molecular gas in ordinary star-forming galaxies (\(\sim10^9~\mathrm{yr}\)), implies that J2054-0005 is undergoing an episode of rapid quenching of star formation. The depletion time is shorter by an order of magnitude compared to that in nearby ULIRGs where OH outflows have been detected \citep{GA17}. On the other hand, since molecular gas is being evacuated from the galactic center region, the outflow is also quenching the gas reservoir available to feed the SMBH (\(M_\mathrm{BH}\approx2\times10^{9}~M_\odot\); \citealt{Far22}), thereby limiting its growth.

The result has important implications for the evolution of the central stellar component and its coevolution with the SMBH, which is believed to be the origin of the relation between the SMBH mass and bulge velocity dispersion found in a large sample of galaxies from the local Universe to high redshifts (e.g., \citealt{MQT05,KH13,Izu19}). The short depletion timescales inferred from this work, as well as other dynamical properties, agree with recent predictions from models of massive galaxy formation at \(z>6\) \citep{Lap18,Pan19}. The models predict that the quasar outflow phase is accompanied by significant stellar component increase up to within \(\sim30~\mathrm{Myr}\), and support the making of quiescent galaxies recently observed at redshifts \(z\approx3\mathrm{-}5\) \citep{Car23}. Our result and other recent OH and OH\(^{+}\) observations \citep{Sha22,But23} indicate that cool gas outflows may be common in quasars at \(z>6\), but further observations of larger samples are needed to investigate their statistical properties.

%%%%%
\subsection{\(\dot{M}_\mathrm{out}\mathrm{-}L_\mathrm{IR}\) relation}

\begin{figure}[ht!]
\epsscale{1.2}
\plotone{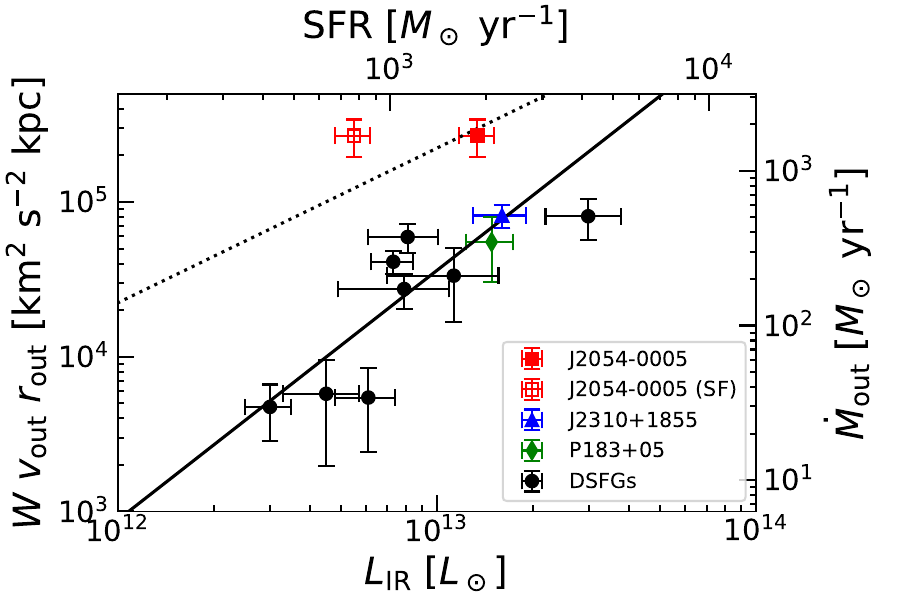}
\caption{Mass outflow rate vs. total IR luminosity (\(L_\mathrm{IR}\)). The ordinate on the left shows the product of measured quantities (equivalent width, outflow velocity, and radius) that is proportional to \(\dot{M}_\mathrm{out}\) in the expanding-shell model; \(\dot{M}_\mathrm{out}=6.28\times10^{-3}Wv_\mathrm{out}r_\mathrm{out}\), where \(W\) is in km s\(^{-1}\), \(v_\mathrm{out}\) is in km s\(^{-1}\), and \(r_\mathrm{out}\) is in kpc. The black circles are dusty star-forming galaxies (DSFGs) at \(z=4\mathrm{-}5\) \citep{Spi20a}. The data of J2310+1855 and P183+05 are taken from \citet{But23}, and \(W\) was calculated using Equation (\ref{eq:eqw}). The ordinate on the right is \(\dot{M}_\mathrm{out}\) calculated using Equations (\ref{eq:cmd}) and (\ref{eq:mor}). The dotted line is \(\eta=1\) and the solid line is the fit. The open square symbol (not included in the fit) shows J2054-0005 corrected for the AGN contribution. The data points for J2310+1855 and P183+05 are not AGN-corrected.\label{fig:Wvr-LIR}}
\end{figure}

Figure \ref{fig:Wvr-LIR} shows a comparison of mass outflow rates and the total IR luminosity (\(L_\mathrm{IR}\)) in high-\(z\) quasars and dusty star-forming galaxies (DSFGs) with OH outflow detections. Since the uncertainty of the mass outflow rate is dominated by the OH abundance, we plot the product \(Wv_\mathrm{out}r_\mathrm{out}\) in addition to \(\dot{M}_\mathrm{out}\), because it is the directly measured quantity that is proportional to \(\dot{M}_\mathrm{out}\) in the expanding-shell model under LTE, whereas \(\dot{M}_\mathrm{out}\) depends on OH abundance, \(f\), and \(T_\mathrm{ex}\). Here, \(W\) is the equivalent width of the outflow component, \(v_\mathrm{out}\) is the center velocity of the absorption line, and \(r_\mathrm{out}\) is the adopted radius of the OH outflow (Section \ref{subsec:mas}). For all sources, we assumed that the radius is equal to the size of the continuum-emitting region in the host galaxy from which the OH spectrum was extracted. This is \(r_\mathrm{out}\approx2.0~\mathrm{kpc}\) for J2054-0005, \(r_\mathrm{out}\approx2.5~\mathrm{kpc}\) for J2310+1855, and \(r_\mathrm{out}\approx4.0~\mathrm{kpc}\) for P183+05, based on Figure 2 in \citet{But23}. For the DSFGs, the outflow radius is assumed to be the continuum size \(r_\mathrm{out}=r_\mathrm{cont}\) from Table 2 in \citet{Spi20a}. \(W\) was calculated using Equation (\ref{eq:eqw}), and we assumed \(T_\mathrm{ex}=50~\mathrm{K}\) and \([\mathrm{OH}]/[\mathrm{H}_2]=5\times10^{-7}\) for all sources in calculating \(\dot{M}_\mathrm{out}\). The plot shows that there is a positive relation between the mass outflow rate with \(L_\mathrm{IR}\), and the power law is \(\log{y}=k\log{x}+m\), where \(k=1.61\pm0.40\) and \(m=-16.4\pm5.2\), although the scatter is large (\(R^2=0.64\)).

\begin{figure}[ht!]
\epsscale{1.15}
\plotone{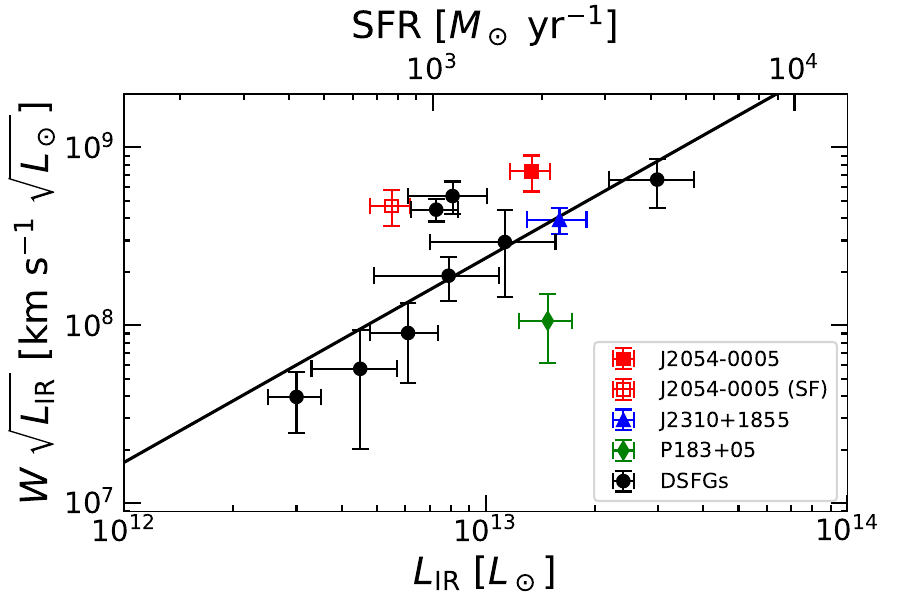}
\caption{\(W\sqrt{L_\mathrm{IR}}\) and total infrared luminosity (\(L_\mathrm{IR}\)). The equivalent width \(W\) is the outflow component obtained from fitting the absorption feature. The solid line is the fitted relation. The open square symbol (not included in the fit) shows J2054-0005 corrected for the AGN contribution. The data points for J2310+1855 and P183+05 are not AGN-corrected.\label{fig:sqrt}}
\end{figure}

As discussed in \citet{HC20} and \citet{Spi20b}, the correlation can also be found if the mass outflow rate is set to be proportional to \(\dot{M}_\mathrm{out}\propto W\sqrt{L_\mathrm{IR}}\). This relation circumvents the uncertainty of the quantities such as the outflow radius and OH optical depth. We reproduce a similar relation in Figure \ref{fig:sqrt}, although \(W\) here is the equivalent width of the outflow (absorption) component instead of \(W_{v<200}\) as in their works. The relation is nearly linear (plotted as solid line) with \(k=1.13\pm0.36\) and intercept \(m=-6.3\pm4.7\) (\(R^2=0.52\)).

The above analysis yields a positive relation between \(\dot{M}_\mathrm{out}\) and \(L_\mathrm{IR}\), though it should be noted that \(L_\mathrm{IR}\) gives an upper limit to the SFR if the AGN fraction is not subtracted, and the fraction may differ among the sources. We have corrected the SFR by subtracting the fractional AGN contribution to dust heating for J2054-0005. Although the sample of quasars is small, Figure \ref{fig:Wvr-LIR} suggests that the \(\dot{M}_\mathrm{out}-L_\mathrm{IR}\) relation may be less clear when such correction is made. The outflows may be driven by the combined contribution of the starburst and AGN (e.g., \citealt{Gow18}), which is responsible for the positive \(\dot{M}_\mathrm{out}\mathrm{-}L_\mathrm{IR}\) relation and the scatter may be caused by nonuniform outflow geometry, projection effects, and different covering factors. On the other hand, \citet{But23} suggest that if a quasar is unobscured, it has already cleared the material near the nucleus, thereby leaving a less energetic outflow due to inefficient coupling with the surrounding ISM material. Since the AGN in J2054-0005 appears to be Type 1, i.e., unobscured (Section \ref{subsec:mor}; \citealt{Jia08}) but the outflow is significantly more powerful compared to those in J2310+1855 and P183+05 (also Type 1), the obscuration effects do not seem to play a major role in outflow energetics.

%%%%%
\subsection{Gas escaping into the IGM}\label{subsec:esc}

Previous observations have revealed the presence of atomic and molecular gas in the halos and circumgalactic medium of high-redshift galaxies (e.g., \citealt{Emo16,Tum17,Fuj20,Cic21,Sch23}). This suggests that the gas was ejected from host galaxies by powerful outflows with terminal velocities that may even exceed the escape velocity. Here, we make a simple analysis to investigate whether the molecular outflow in J2054-0005 is fast enough to transport OH gas into the IGM.

The dynamical mass of J2054-0005, derived from [\ion{C}{2}] \(158~\micron\) data for a disk inclination angle of \(24\arcdeg\), is estimated to be \(M_\mathrm{dyn}\approx7.2\times10^{10}M_\sun\) \citep{Wan13} within the [\ion{C}{2}]-emitting region of radius \(R\approx1~\mathrm{kpc}\). Assuming a spherically symmetric mass distribution, this yields an escape velocity of \(v_\mathrm{esc}(R)\approx\sqrt{2GM_\mathrm{dyn}/R}\approx780~\mathrm{km~s^{-1}}\) at \(R\). The estimate corresponds to a rotational velocity of \(560~\mathrm{km~s^{-1}}\), implying a very massive host galaxy. The escape velocity is similar to the velocities reported for dusty star-forming galaxies at \(z=4\mathrm{-}5\) \citep{Spi20b}. Since the obtained value is comparable to the mean velocity of the outflow along the line of sight, a significant fraction (up to \(\sim50\%\)) of the outflowing molecular gas may be able to escape the gravitational potential well and inject metals and dust into the IGM. This is in agreement with recent observations of enriched gas in the circumgalactic medium at \(z\sim6\) \citep{Wu21}.

%%%%%
\subsection{[\ion{O}{3}]88/[\ion{C}{2}]158 luminosity ratio}\label{subsec:fsl}

Recently, the luminosity ratio of [\ion{O}{3}] 88 \(\micron\) to [\ion{C}{2}] 158 \(\micron\) has drawn attention as it is found to be higher at high-redshift compared to local star-forming galaxies (e.g., \citealt{Ino16,Lap19,Has19,Has19b,Pal19,Tam19,Ara20,Bak20,Carn20,Har20,Lup20,Val21,Kat22,Sug22,Wit22,Ren23}) and local dwarf galaxies \citep{Ura23}. One of the scenarios proposed to explain the observations is the possibility of outflows affecting the covering factor of photodissociation regions (PDRs) \citep{Har20}. In sources with powerful outflows, low-ionization PDRs traced by [\ion{C}{2}] 158 \(\micron\) may be cleared so that their covering factor is decreased relative to that of \ion{H}{2} regions traced by [\ion{O}{3}] 88 \(\micron\). If that is the case, we may expect to see more powerful outflows in sources with a high luminosity ratio.

So far, only two reionization-epoch quasars (J2054-0005 and J2310+1855) with OH outflow detections have been detected also in [\ion{C}{2}] 158 \(\micron\) and [\ion{O}{3}] 88 \(\micron\) \citep{Has19,But23}. A comparison of their properties shows the following characteristics. (1) The luminosity ratio \(L_\mathrm{[OIII]}/L_\mathrm{[CII]}\) is \(\sim7\) times larger in J2054-0005 (\(2.1\pm0.4\)) compared to J2310+1855 (\(0.3\pm0.1\)). (2) The mass outflow rate in J2054-0005 is \(\sim3\) times larger than in J2310+1855 (Figure \ref{fig:Wvr-LIR}); the mean outflow velocity is also higher (\(669\pm87~\mathrm{km~s}^{-1}\) in J2054-0005 compared to \(334\pm14~\mathrm{km~s}^{-1}\) in J2310+1855). On the other hand, J2310+1855 has slightly larger total IR luminosity (\(1.9\times10^{13}~L_\odot\)) compared to J2054-0005 (\(1.3\times10^{13}~L_\odot\)). Although we cannot draw conclusions from only two sources, J2054-0005, with a more powerful outflow, also has a significantly higher luminosity ratio, which supports the scenario of a low PDR covering factor. Previous [\ion{C}{2}] 158 \(\micron\) studies suggest that the inclination angle of a rotating host galaxy in J2054-0005 is relatively low (\(\approx24\arcdeg\); \citealt{Wan13}). If that is the case, and if the outflow is predominantly propagating perpendicular to a rotating disk, it is close to the line of sight, hence the observed velocity is higher compared to that in J2310+1855.

%%%%%
\subsection{OH emission: highly excited molecular gas}\label{subsec:ohe}

The OH \(119~\micron\) line has been detected in emission toward a number of nearby AGNs \citep{Spi05,Vei13}. Recently, \citet{But23} reported a detection of OH emission in one quasar at \(z\approx6\). However, the line has been detected only in absorption toward a sample of dusty star-forming galaxies at \(z=4\mathrm{-}5\) \citep{Spi20a}.

Based on multi-line OH observations, \citet{Spi05} argue that collisional excitation dominates the population of the \(^{2}\Pi_{3/2}~J=5/2\) level that leads to radiative decay and \(119~\micron\) emission in the active nucleus of the local Seyfert galaxy NGC 1068. On the other hand, \citet{Vei13} found that the strength of the OH \(119~\micron\) absorption relative to emission is correlated with the \(9.7~\micron\) silicate strength, an indicator of the obscuration of the nucleus, in their ULIRG sample. \citet{Spo13} argue that the OH emission arises from dust-obscured central regions and that, except in two outliers, radiative excitation may be dominant. Since the peak of the spectral energy distribution of dust thermal emission in J2054-0005 is close to \(\lambda_\mathrm{rest}=53~\micron\), the wavelength that corresponds to the energy difference between the levels \(^{2}\Pi_{1/2}~J=3/2\) and \(^{2}\Pi_{3/2}~J=3/2\), absorption of the continuum from dust emission could excite the level which would radiatively decay into the ground state. In that case, it is expected that the \(^{2}\Pi_{1/2}~J=3/2\rightarrow1/2\) line at \(\lambda_\mathrm{rest}=163~\micron\) would also be observed in emission. Given that the \(120~\micron\) continuum emission is spatially extended, and the fact that OH emission is marginally spatially resolved (Section \ref{subsec:res}), it is likely that the excited OH gas is not confined to the compact AGN, but distributed in a broader (\(\sim2~\mathrm{kpc}\)) region. Further multi-line observations are necessary to constrain the excitation mechanism of OH molecules in EoR quasars.

%%%%%%%%%%%%%%%%%
\section{Summary}\label{sec:sum}
%%%%%%%%%%%%%%%%%

We have presented the first ALMA observations of the OH 119 \(\micron\) (\(^2\Pi_{3/2}~J=5/2\mathrm{-}3/2\)) line toward the reionization-epoch quasar J2054-0005 at redshift \(z\approx6.04\) at the resolution of \(0\farcs20\times0\farcs17\). The main findings reported in the paper are summarized below.

\begin{enumerate}

\item{The OH 119.23, 119.44 \(\micron\) doublet line and the \(120~\micron\) continuum are detected toward the quasar. The continuum is detected at high signal-to-noise ratio of \(260\). The OH line exhibits a P-Cygni profile with absorption and emission components.}

\item{We fitted the OH profile with two double-gaussian functions using a least-squares fitting tool. The fits reveal a blue-shifted absorption component (peak absorption depth \(\tau\approx0.36\)), that unambiguously reveals as outflowing molecular gas, and emission component at near-systemic velocity. The absorption peak velocity is \(v_\mathrm{cen}=-669\pm87~\mathrm{km~s}^{-1}\), the FWHM line width is \(1052\pm234~\mathrm{km~s}^{-1}\), and the terminal velocity is \(v_{98}=-1574\pm35~\mathrm{km~s}^{-1}\), indicating a fast molecular outflow. This is the first quasar with such high molecular outflow velocity discovered at \(z>6\).}

\item{The mass outflow rate, calculated under LTE approximation, OH abundance \([\mathrm{OH}]/[\mathrm{H}_2]=5\times10^{-7}\), and assuming the geometry of an expanding thin spherical shell with a covering factor \(f=0.3\) and radius \(r_\mathrm{out}=2~\mathrm{kpc}\), is \(\dot{M}_\mathrm{out}\approx1700~M_\odot~\mathrm{yr}^{-1}\). Using an empirical relation from the literature, we obtain \(\dot{M}_\mathrm{out}^\mathrm{emp}\approx1500~M_\odot~\mathrm{yr}^{-1}\).}

\item{The absorption and emission components of the OH line are marginally spatially resolved in the central 2 kpc, suggesting that the outflow extends over this region. Since the critical density and excitation energy for the upper rotational levels of OH are relatively high (\(n_\mathrm{cr}\gtrsim10^9~\mathrm{cm}^{-3}\), \(E/k=120~\mathrm{K}\)), the detection of OH emission implies that molecular gas is highly excited (warm or dense, shocked), possibly by far-IR radiation pumping from dust grains and by collisions with H\(_2\) (e.g., shocks). The OH line is significantly broader compared to [\ion{C}{2}] 158 \(\micron\) in the central 1-kpc region.}

\item{In order to estimate the fractional contribution of AGN to the total infrared luminosity (\(L_\mathrm{IR}\)), we performed SED fitting of the spectrum of J2054-0005 using the code \texttt{CIGALE}. The result yields a total infrared luminosity of \(L_\mathrm{IR}=(1.34\times0.17)\times10^{13}~L_\sun\) and suggests that as much as \(59\%\) of \(L_\mathrm{IR}\) is produced by dust heated by the AGN and not by star formation. The IR-derived SFR corrected for this effect is estimated to be \(\mathrm{SFR}=770\pm180~M_\sun~\mathrm{yr}^{-1}\) (Chabrier IMF).}

\item{The mass outflow rate is comparable to the AGN-corrected star formation rate in the host galaxy (\(\dot{M}_\mathrm{out}/\mathrm{SFR}\sim2\)); it is higher compared to other two quasars with OH detections at \(z>6\) and among the highest at high redshift. At the current mass loss rate, molecular gas is expected to be depleted after only \(t_\mathrm{dep}\approx(2\mathrm{-}4)\times10^7~\mathrm{yr}\), implying rapid quenching of star formation. The dynamical age of the outflow is \(t_\mathrm{out}=r_\mathrm{out}/v_\mathrm{out}\approx3\times10^6~\mathrm{yr}\).}

\item{An analysis of the outflow energetics and terminal velocity indicates that the outflow in J2054-0005 may be powered by the combined effects of the AGN and star formation. This is supported by the fact that we find a positive correlation between \(\dot{M}_\mathrm{out}\) and total luminosity \(L_\mathrm{IR}\) using a sample of 8 dusty star-forming galaxies at \(z=4\mathrm{-}5\) and 3 quasars at \(z>6\), that the outflow velocity in J2054-0005 is relatively high compared to the majority of star-forming galaxies at high \(z\), and that as much as \(59\%\) of \(L_\mathrm{IR}\) is produced by the AGN.}

\item{The mean outflow velocity is comparable to the estimated escape velocity. This implies that as much as \(\sim50\%\) of the outflowing molecular gas may be able to escape from the host galaxy and enrich the intergalactic medium with heavy elements.}

\item{We report the discovery of a companion at a projected separation of \(2\farcs4\). This source is detected only in continuum at the significance of \(8.9~\sigma\).}

\end{enumerate}

\begin{acknowledgments}
The authors thank the referee for comments and suggestions that helped us improve the manuscript. This paper makes use of the following ALMA data: ADS/JAO.ALMA\#2021.1.01305.S, ADS/JAO.ALMA\#2019.1.00672.S. ALMA is a partnership of ESO (representing its member states), NSF (USA) and NINS (Japan), together with NRC (Canada), MOST and ASIAA (Taiwan), and KASI (Republic of Korea), in cooperation with the Republic of Chile. The Joint ALMA Observatory is operated by ESO, AUI/NRAO and NAOJ. DS was supported by the ALMA Japan Research Grant of NAOJ ALMA Project, NAOJ-ALMA-294. TH was supported by Leading Initiative for Excellent Young Researchers, MEXT, Japan (HJH02007) and by JSPS KAKENHI Grant Number 22H01258. DD acknowledges support from the National Science Center (NCN) grant SONATA (UMO-2020/39/D/ST9/00720). This study is supported by JSPS KAKENHI Grant Number 17H06130, 20H01951, 22H04939, and NAOJ ALMA Scientific Research Grant Number 2018-09B. This work was supported by JST SPRING, Grant Number JPMJSP2119.
\end{acknowledgments}

\software{CASA \citep{TCT22}, SciPy \citep{Vir20}}

\end{document}